\newcommand*{\ATLASLATEXPATH}{latex/}

\newcommand{\Zgjj}{\ensuremath{Z \gamma jj}\xspace}

\newcommand{\llgjj}{\ensuremath{\ell^{+} \ell^{-} \gamma jj}\xspace}

\newcommand{\cZg}{\ensuremath{\zeta_{Z\gamma}}\xspace}
\newcommand{\pTbalance}{\ensuremath{p_{\mathrm{T}}^{\mathrm{balance}}}\xspace}

\newcommand{\OO}{\mathcal{O}}

\def\alpgen{\textsc{Alpgen} }
\def\powheg{\textsc{Powheg} }
\def\sherpa{\textsc{Sherpa} }
\def\vbfnlo{\textsc{Vbfnlo}}
\def\POWHEGBOX{\textsc{Powheg-Box}\xspace}

\pdfoutput=1

\documentclass[cernpreprint,texlive=2016,txfonts,UKenglish]{\ATLASLATEXPATH atlasdoc}

\usepackage[biblatex=true,subfigure]{\ATLASLATEXPATH atlaspackage}

\usepackage{\ATLASLATEXPATH atlasbiblatex}

\usepackage{\ATLASLATEXPATH atlascontribute}

\usepackage{\ATLASLATEXPATH atlasphysics}

\graphicspath{{logos/}{figures/}}

\usepackage{PaperVyVBS-defs}
\usepackage{multirow}
\usepackage{xspace}
\usepackage{textcomp}
\usepackage{setspace}
\usepackage{placeins}
\usepackage{a4wide}

\AtlasTitle{Studies of  $Z\gamma$  production in
association with a high-mass dijet system in $pp$ collisions at
$\sqrt{s}=8$~\TeV with the ATLAS detector}

\author{The ATLAS Collaboration}

\AtlasRefCode{STDM-2015-21}

\PreprintIdNumber{CERN-EP-2017-046}

 \AtlasJournal{JHEP}

\AtlasAbstract{

The production of a $Z$ boson and a photon in association with a
high-mass dijet system is studied using 20.2 fb$^{-1}$ of
proton–proton collision data at a centre-of-mass energy of $\sqrt{s} = 8$~\TeV{}
recorded with the ATLAS detector in 2012 at the Large Hadron Collider.
Final states with a photon and a Z boson decaying into a pair of either electrons,
muons, or neutrinos are analysed. Electroweak and total $pp \to Z\gamma j j$
cross-sections are extracted in two fiducial regions with different sensitivities to
electroweak production processes. Quartic couplings of vector bosons are studied
in regions of phase space with an enhanced contribution from pure electroweak
production, sensitive to vector-boson scattering processes $VV \to Z\gamma$.
No deviations from  Standard Model predictions are observed and constraints are placed on
anomalous couplings parameterized by higher-dimensional operators using effective field
theory.

}

\AtlasCoverSupportingNote{~~~~~~~~~~~~~~~~~~~~~~ATL-COM-PHYS-2015-1433}{https://cds.cern.ch/record/2109528}
\AtlasCoverSupportingNote{~~~~~~~~~~~~~~~~~~~~~~Changes between drafts 2.0 and 3.0}{https://indico.cern.ch/event/624769/contributions/2522577/attachments/1442266/2223913/Summary_ZyVBS_AQGCchanges.pdf}

 \AtlasCoverCommentsDeadline{ April 19th, 2017}

 \AtlasCoverAnalysisTeam{Andrea Bocci, Hal Evans, Alfred Goshaw, Shih-Chieh Hsu, Ashutosh Kotwal, Dimitriy Krasnopevtsev, Shu Li, Zhijun Liang, Narei Lorenzo Martinez, Christopher Meyer, Muhammad Firdaus Mohd Soberi, John Penwell, Anatoli Romaniouk, Evgeny Soldatov, Ben Tannenwald, Rick Van Kooten }

 \AtlasCoverEdBoardMember{Marcello Fanti (chair), Tulay Cuhadar, Hulin Wang}

 \AtlasCoverEgroupEditors{atlas-stdm-2015-21-editors}

 \AtlasCoverEgroupEdBoard{atlas-stdm-2015-21-editorial-board}

\hypersetup{pdftitle={ATLAS draft},pdfauthor={The ATLAS Collaboration}}

\begin{document}

\maketitle

\section{Introduction}
\label{sec:introduction}

The scattering of two vector bosons, $V V \to V V$ with $V$ =
$W$/$Z$/$\gamma$, is a key process for probing the ${\mathrm{SU(2)}}_\mathrm{L}\times {\mathrm{U(1)}}_\mathrm{Y}$
 gauge symmetry of the electroweak theory that determines the
self-couplings of the vector bosons. In particular, it is important to
independently test the triple and quartic gauge-boson coupling
strengths (TGCs and QGCs), since new phenomena could generate
additional contributions to QGCs with respect to the Standard Model
(SM) predictions~\cite{bib:eboli, bib:eboli-dim8, bib:snowmass_ewk,
  bib:eft}, while not significantly affecting the TGCs
~\cite{bib:artz}.

Experimental information about QGCs is still limited. At the Large Hadron Collider (LHC), it can be deduced 
from studies of processes with either three bosons in the final
state~\cite{bib:CMSWVa,Aad:2015uqa,bib:CMSVyy,bib:zg2012,bib:www} or involving
pure electroweak production of heavy
dibosons~\cite{bib:CMSWWaa,bib:ggWW,bib:ssWW2,bib:WWhad}.
In particular, the CMS experiment recently studied
$Z\gamma$~\cite{bib:cmsZy}
and $W\gamma$~\cite{bib:cmsWy} electroweak production and used these results to set limits on
anomalous QGCs.  
Final states involving photons have higher rates than those 
involving only $W$ and $Z$ bosons identified via their leptonic decay modes.

\begin{figure}[h!]
  \begin{center}
  \includegraphics[width=0.33\columnwidth]{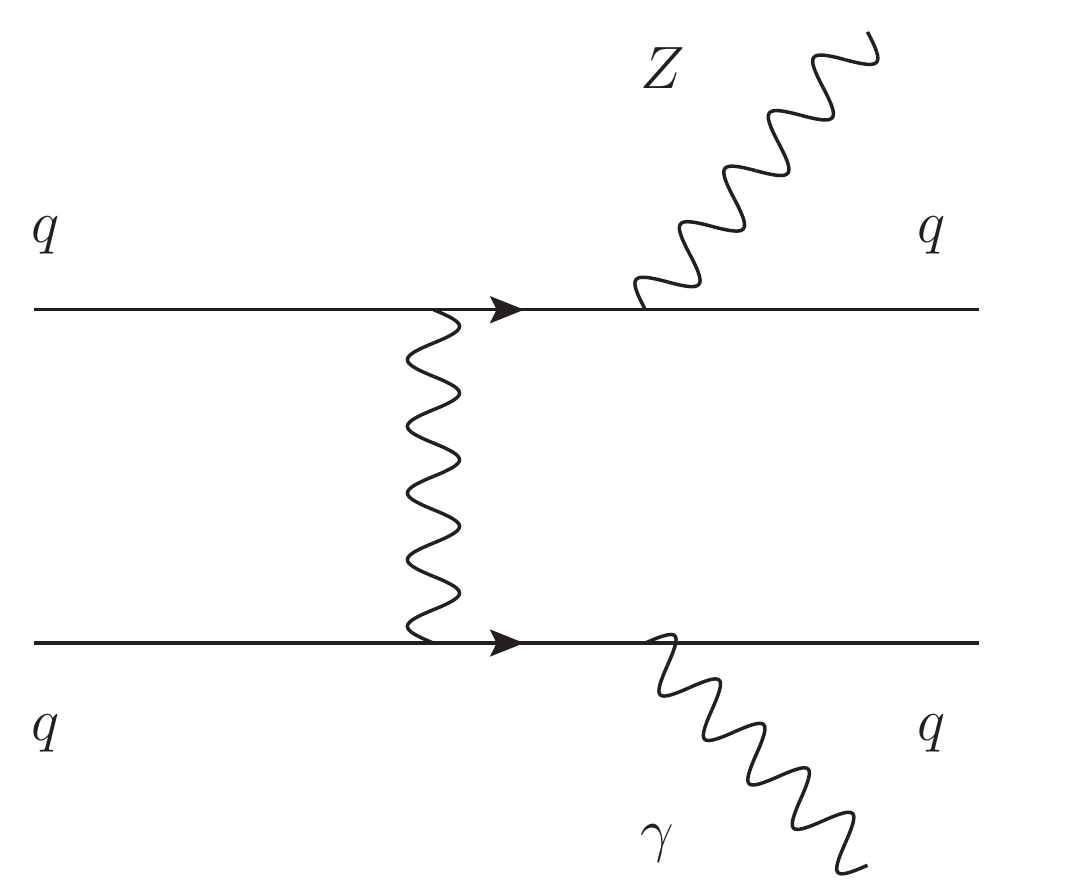}
  \hspace{2cm}
  \includegraphics[width=0.33\columnwidth]{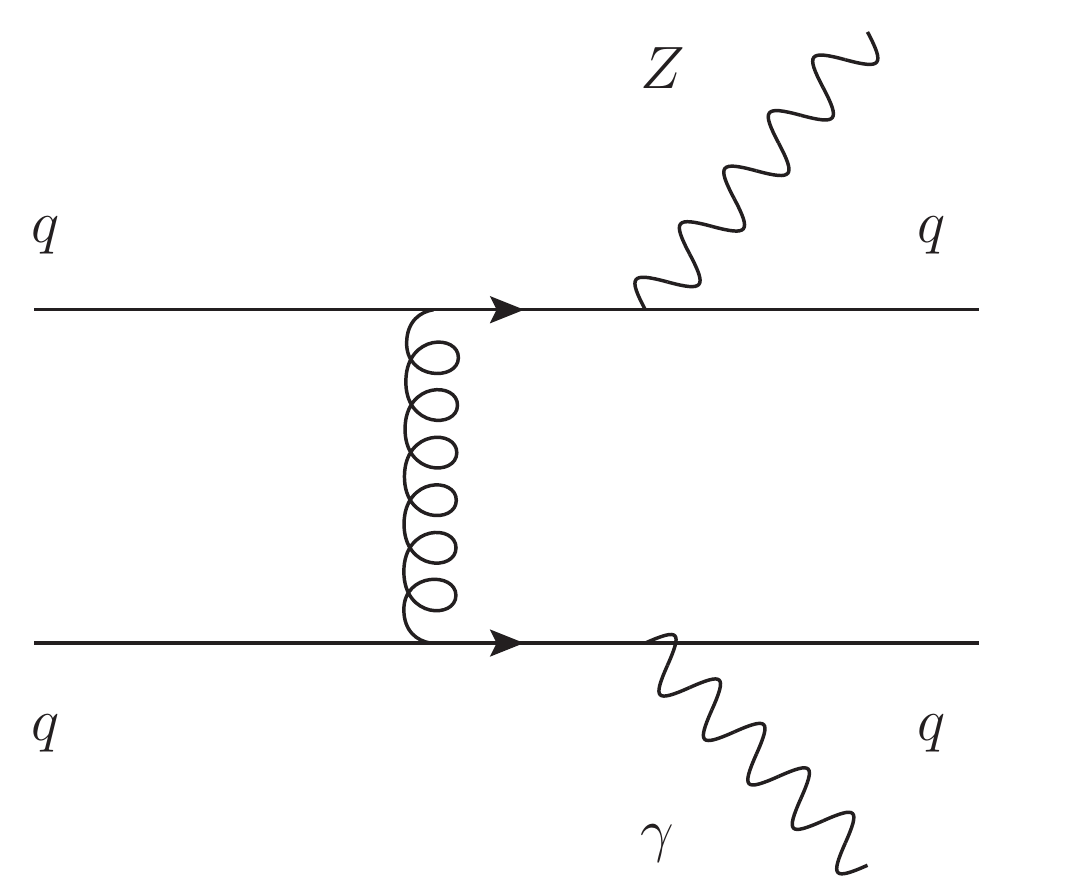} \\
  \includegraphics[width=0.33\columnwidth]{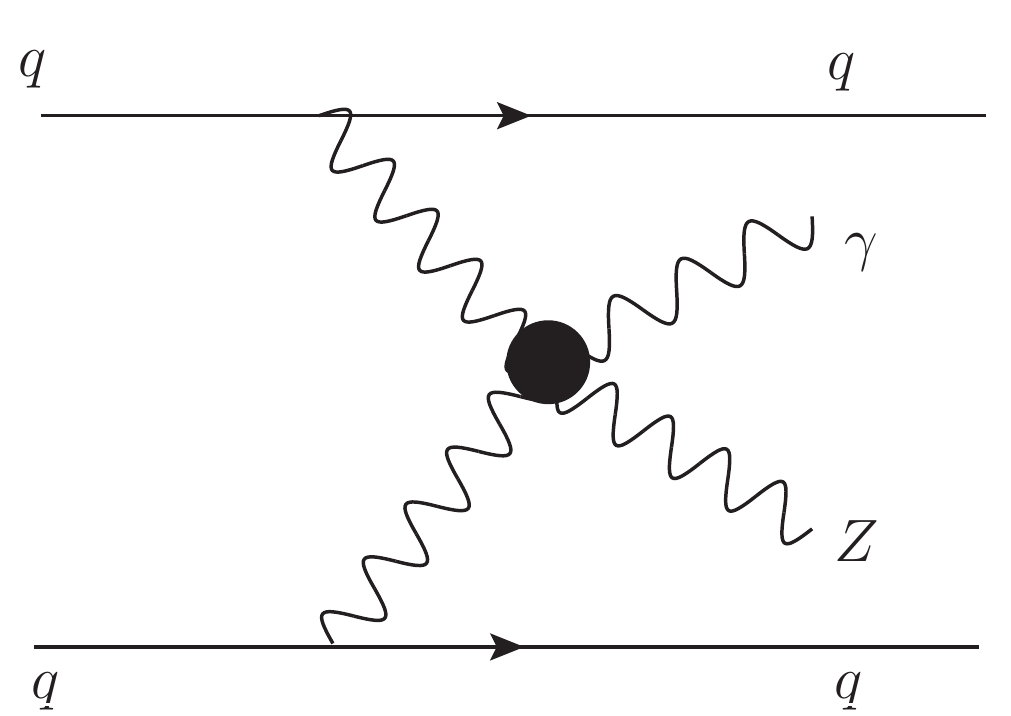}
  \hspace{2cm}
  \includegraphics[width=0.33\columnwidth]{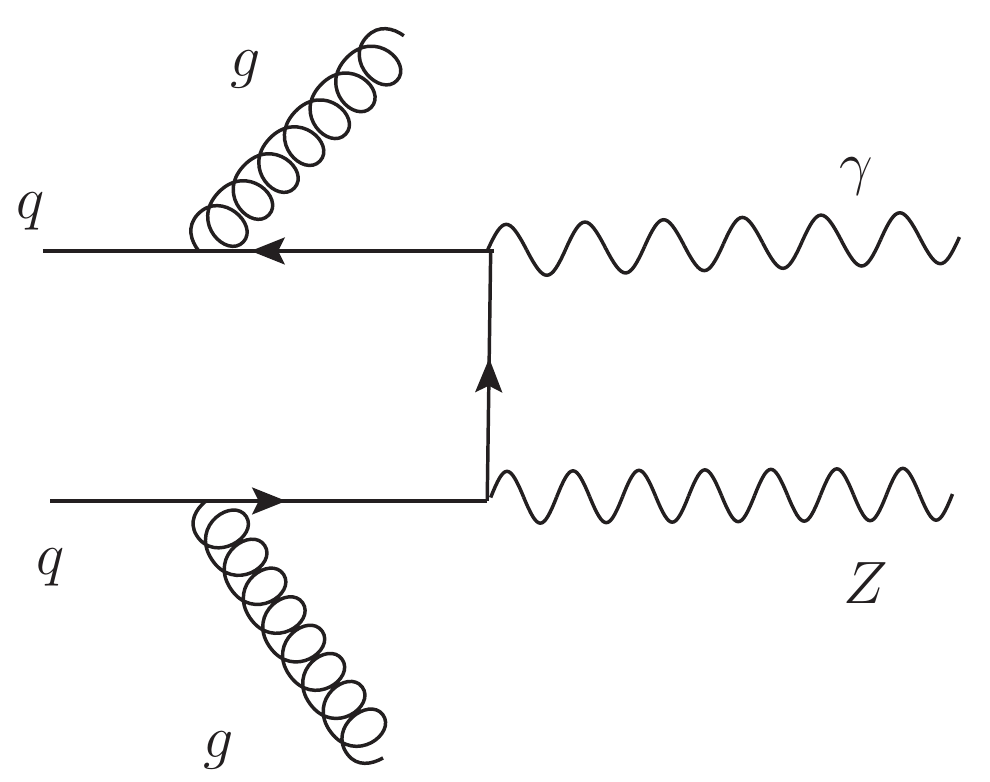}
  \caption{Feynman diagrams of electroweak $Z\gamma jj$ production
    involving VBS subprocesses (bottom left) or non-VBS subprocesses (top left) and of QCD $Z\gamma jj$ production with gluon exchange (top right) or radiation (bottom right).} 
  \label{fig:fey_dia}
  \end{center}
\end{figure}

The \Zgjj electroweak (EWK) production  ($qq \to qq Z\gamma$) ---
where $j$ represents a jet and $q$ a quark --- contains processes with
fourth-order electroweak coupling $\OO(\alpha_\mathrm{em}^4)$. These include
vector-boson scattering (VBS) as well as non-VBS diagrams, e.g. when
the $Z$ boson and the photon are radiated off the initial- or final- state
quarks (Figure~\ref{fig:fey_dia}, left). The VBS processes do not
respect the electroweak gauge symmetry when taken in isolation and
cannot be studied separately from other electroweak processes, due to
large interference effects.

The same \Zgjj final state can be produced by QCD-mediated processes
--- in the following simply called ``QCD production'' --- with
second-order electroweak coupling and second-order strong coupling
$\OO(\alpha_{\mathrm{em}}^2\alpha_{\mathrm{s}}^2)$  (Figure~\ref{fig:fey_dia}, right). Such
processes can involve radiated gluons in the initial and/or final
state as well as quark scattering processes mediated
by gluons.
According to the SM, a small constructive interference occurs between production of
QCD and EWK quark scattering.

Experimentally, \Zgjj\ EWK processes are characterized by the
production of two energetic hadronic jets with wide rapidity
separation and large dijet invariant mass~\cite{bj1993}. The
vector-boson pair is typically produced more centrally than in non-EWK processes. These
kinematic properties are exploited to select a phase-space region where the
electroweak production is enhanced with respect to the QCD-mediated processes. 

Previous measurements of inclusive and differential cross-sections of
 $Z\gamma$ production in proton--proton collisions at the
centre-of-mass energy of 8~\TeV performed by the ATLAS
experiment~\cite{bib:zg2012} show good agreement (within 6\%)
between data and next-to-next-to-leading-order (NNLO) predictions.

A study of \Zgjj\ EWK production is made in a search region
defined to maximize its expected significance.
This corresponds to a phase-space region with a high-mass dijet system, a high-energy
photon, and a $Z$ boson decaying into a pair of electrons or muons --- in the following referred
to as the "charged-lepton channel". 
The contribution from QCD production is
constrained with data from a background-enriched control region,
corresponding to events with lower dijet invariant mass.
A fiducial cross-section is extracted and compared to
SM predictions.
The total \Zgjj\ (EWK+QCD) production cross-section in both the search and control
fiducial regions is also measured.

The \Zgjj EWK production is also studied in events with
high-transverse-energy ($E_{\mathrm{T}}$) photons, where an
enhancement of the VBS cross-section is typically predicted by
theories beyond the Standard Model. In this phase space region, a
search for anomalies in the quartic $WWZ\gamma$ coupling along with
the presence of $ZZZ\gamma$, $ZZ\gamma \gamma$ and
$Z\gamma \gamma \gamma$ couplings --- forbidden in the SM at tree
level --- is performed. Extensions of the SM Lagrangian are parameterized by
higher-order operators in an effective field theory that modifies the
boson couplings. The sensitivity of the search is improved by also
considering events with the $Z$ boson decaying into neutrinos. The
phase-space region optimized for anomalous QGC (aQGC) sensitivity is
defined differently for the charged-lepton and neutrino channels. For
the latter, since the $Z$ boson is experimentally invisible, extra
requirements are included to suppress the relatively large backgrounds
present in this channel.

The paper is organized as follows. A brief description of the ATLAS
detector and the data set used in this analysis is given in
Section~\ref{sec:detector}. The simulation of the contributing
processes is summarized in Section~\ref{sec:simulation}, while event
reconstruction and selection is detailed in
Section~\ref{sec:selection}. The determination of the backgrounds and
event yields are discussed in Section~\ref{sec:background}. The
extraction of the cross-sections are described in
Section~\ref{sec:xsec}. Finally, a search for aQGCs 
using events with high-$E_{\mathrm{T}}$ photons is presented in
Section~\ref{sec:aqgc}.  Conclusions are drawn in
Section~\ref{sec:conclusions}.

\section{ATLAS detector and data}
\label{sec:detector}

The ATLAS experiment~\cite{Aad:2008zzm} at the LHC is a multipurpose
particle detector with a forward--backward symmetric cylindrical
geometry and an almost $4\pi$ coverage in solid angle.\footnote{\AtlasCoordFootnote} It consists of 
a tracking system called the inner
detector (ID) surrounded by a thin superconducting solenoid providing a
\SI{2}{\tesla} axial magnetic field, electromagnetic and hadronic
calorimeters, and a muon spectrometer (MS).  The 
ID covers the pseudorapidity range $|\eta| < 2.5$.  It consists of
silicon pixel, silicon microstrip, and transition radiation tracking
detectors.  Within the region of $|\eta|< 3.2$, electromagnetic (EM) calorimetry is
provided by high-granularity lead/liquid-argon (LAr) sampling
calorimeters, with an additional thin LAr presampler covering $|\eta|
< 1.8$ to correct for energy loss in material upstream of the
calorimeters.  A hadronic (steel/scintillator-tile) calorimeter covers
the central pseudorapidity range ($|\eta| < 1.7$).  The endcap and
forward regions are instrumented with LAr calorimeters for both the EM
and hadronic energy measurements up to $|\eta| = 4.9$.  The MS surrounds the calorimeters and is based on three large
air-core toroidal superconducting magnets with eight coils each.  It
includes a system of precision tracking chambers and fast detectors
for triggering.  A three-level trigger system is used to select
events.  The first-level trigger is implemented in hardware and uses a
subset of the detector information to reduce the accepted rate to at
most \SI{75}{\kilo\hertz}.  This is followed by two software-based systems, called the
high-level triggers, that together reduce the accepted event rate to
\SI{400}{\hertz} on average, depending on the data-taking conditions.

The data set used in this analysis was obtained from proton--proton
collisions recorded in 2012 by the ATLAS detector, when the LHC
operated at $\sqrt{s} = 8~\TeV$.  The integrated luminosity of the
data set used in this measurement is 20.2~\ifb with an uncertainty of
1.9\%~\cite{Aaboud:2016hhf}.

In the charged-lepton channel analysis, events are selected online
by requiring the presence of either an isolated electron or muon
candidate with a minimum transverse momentum ($p_{\mathrm{T}}$) of
24~\GeV, or a pair of isolated electron candidates with
$p_{\mathrm{T}}>$ 12~\GeV, or a pair of isolated muon candidates satisfying 
$p_{\mathrm{T}} >$~18~\GeV and $p_{\mathrm{T}}>$~8~\GeV for the
leading and subleading muons. Trigger efficiencies are included
in the overall reconstruction efficiency, and the uncertainties in the
efficiency of these trigger selections were estimated using control
samples in data and amount to 0.2\% and 0.5\% in the $e^+e^-\gamma jj$
and $\mu^+\mu^-\gamma jj$ channels, respectively.

In the neutrino channel, the events are selected online by requiring 
a photon candidate with 
$E_{\mathrm{T}} >$~40~\GeV and missing transverse
momentum greater than 60~\GeV. Trigger efficiencies are included
in the overall reconstruction efficiency, and the uncertainties in the efficiency of
these trigger selections were estimated using control samples in data
and amount to 2.0\%.

\section{Simulated samples and theory predictions}
\label{sec:simulation}

Monte Carlo (MC) event samples, using a full simulation~\cite{ATLASsim}
of the ATLAS detector by \GEANT4~\cite{geant4},
are used to model the data, including contributions from the SM signal and expected backgrounds.

The individual \Zgjj\ EWK and the $Z\gamma$ QCD (with up to three
additional final-state partons) processes are modelled at leading
order (LO) with the \sherpa event generator v1.4.5~\cite{bib:sherpa}. The EWK--QCD
interference contribution is predicted from \madgraph ~\cite{Madgraph} to be less than
10\% of the EWK cross-section in the search region --- invariant mass
of the two leading jets, $m_{jj}$, greater than $500$~\GeV --- with a
decreasing trend as a function of $m_{jj}$.  This interference is
treated as an uncertainty in the measurements, as discussed in
Section~\ref{sec:XSewk}.  Major background processes, such as
$Z$+jets, $\gamma$+jets, $W$+jets, $WW$+jets, and $WZ$+jets are also
modelled by the \sherpa event generator. These include up to five
additional final-state partons at LO for the $V+$jets processes and up
to three additional partons at LO for the $VV+$jets processes. All the
{\sherpa} samples include parton showering (with the CKKW matching
scheme~\cite{Catani:2001cc,Krauss:2002up} scale set to 20~\GeV), and
fragmentation processes along with simulation of the underlying event.
They are generated using the {\sc CT10}~\cite{Lai:2010vv} parton
distribution function (PDF) set.

Uncertainties in the \sherpa modelling of the \Zgjj processes are
estimated using 68\% confidence-level PDF uncertainties, 
independent variations of renormalization and factorization scales by a
factor of two and variations of the choice of CKKW scale (from 15 to
30~\GeV).

Production of $t\bar{t}$ pairs is modelled by \MCatNLO v4.06~\cite{MCatNLO1,MCatNLO2}, 
interfaced to \HERWIG v6.520.2 for
parton showering and fragmentation, and to \Jimmy v4.31.3~\cite{jimmy}
for underlying-event simulation. The $t\bar{t}\gamma$ production is
modelled with \MGMCatNLO \ v5.2.1.2~\cite{Madgraph} and the {\sc CTEQ6L1}~\cite{cteq} 
PDF set, with parton showering, hadronization,
and the underlying event modelled by \pythia v8.183~\cite{Sjostrand:2007gs}. 
The cross-section is computed at next-to-leading-order (NLO) according to Ref.~\cite{Melnikov:2011ta}.

Some of the systematic uncertainties of the $Z$+jets background
estimate, which is extracted from data, are estimated using $Z$+jets NLO \POWHEGBOX v1.0 and LO
\alpgen v2.1.4 (with up to five additional final-state partons) generated
events.  These samples are interfaced with \pythia v8.175 and \HERWIG v6.520.2 +
\Jimmy v4.31.3 respectively for the modelling of the parton shower,
hadronization and underlying event.

Multiple proton--proton collisions (pile-up), corresponding to the conditions observed 
during the 2012 run, are added to each MC sample. This pile-up is simulated 
using \pythia v8.165~\cite{Sjostrand:2007gs} with parameter values set according to
the A2 tune~\cite{ATL-PHYS-PUB-2012-003} and the MSTW2008LO PDF set~\cite{Martin:2009iq}. 
MC events are then reweighted so that the pile-up 
conditions in the simulation match those observed in the data.

The SM cross-section predictions for both the \Zgjj EWK and QCD
processes with exactly two additional final-state partons are
calculated at NLO precision in $\alpha_{\mathrm{s}}$  using the \vbfnlo\
event generator v2.7.1~\cite{bib:vbfnlo1,bib:vbfnlo2,bib:vbfnlo3}. All spin
correlations and finite-width effects are included in the
matrix-element calculation, and for EWK production all resonant and
non-resonant $t$-channel exchange contributions giving rise to a
specific leptonic final state are considered.
The {\sc CT10} PDF set is used for both EWK and QCD production as well as for the underlying-event generation and tune.
These samples are generated separately
(i.e.  the interference between EWK and QCD is not taken into account in the modelling). A photon isolation
requirement to remove the contributions from partons collinear to the
photon is also included in the calculation. The uncertainty in these
predictions due to the PDF and the choice of
renormalization/factorization scale ranges from 4\% to 10\% depending
on the processes and phase-space regions.

\section{Event reconstruction and selection}
\label{sec:selection}

\subsection{Event reconstruction}
\label{sec:obj}

Events are required to have a reconstructed primary vertex 
formed by at least two tracks with $p_{\mathrm{T}} > 400$~\MeV 
and $|\eta|<2.5$. If more than one primary vertex is found, the
one with the largest sum of the $p_{\mathrm{T}}^2$ of the associated tracks is
chosen as the hard-interaction vertex.

Electron candidates, reconstructed by matching an energy deposit in
the calorimeter to a track in the ID, are required to have
$E_{\mathrm{T}}>25$~\GeV and $|\eta|<2.47$. In addition, they must
satisfy a set of ``Loose''~\cite{Aaboud:2016vfy} identification
criteria based on a combination of shower shape information from the
EM calorimeter and tracking information from the ID, corresponding to
an average selection efficiency of about 93\%. The electron tracks are
required to have longitudinal impact parameter smaller than 0.5 mm
with respect to the hard-interaction vertex, and the absolute value of
the transverse impact parameter with respect to the primary vertex
less than six times its measured uncertainty, to reduce semi-leptonic
heavy flavor decay backgrounds. Electron candidates are also required
to be isolated. This is achieved by requiring the sum of the
transverse momenta of ID tracks associated with the primary vertex in
a cone of size $\Delta R=0.3$ around the electron direction, excluding
the electron track, to be less than 10\% of the transverse energy of
the electron candidate itself. Uncertainties in the electron selection
arise from: identification~\cite{Aaboud:2016vfy} and impact parameter
selection variations; changes in the isolation definition; and
uncertainties in the electron energy scale and
resolution~\cite{Aad:2012elid}.

Muon candidates are reconstructed by combining tracks in the ID with
tracks in the MS and are required to have $p_{\mathrm{T}}>25$~\GeV and
$|\eta|<2.5$. The ID tracks associated with these muons must satisfy
several quality selection criteria~\cite{muon}. The same requirement
on the longitudinal impact parameter as for the electron track is also imposed
on the combined muon track. The overall selection efficiency of
the muon identification is about 97\%. Muon candidates are required to
be isolated using the same criteria as for electrons, but using a cone
of size $\Delta R=0.2$. 
Uncertainties in the muon selection are derived from uncertainties in the muon momentum scale and resolution~\cite{muon},
and by varying the selection criteria on the muon track quality, impact parameter or isolation.

Photon reconstruction and identification criteria are based on
the expected shapes of showers developing in the electromagnetic calorimeter, 
 as described in Ref.~\cite{Aaboud:2016yuq}. Photons must be within
the fiducial volume of the central calorimeter ($|\eta|<2.37$) and
outside the transition region between the barrel and endcap
calorimeters ($1.37<|\eta|<1.52$). The sum of the transverse energies
of topological clusters reconstructed in the electromagnetic and
hadronic calorimeters in a cone of size $\Delta R=0.4$ around the photon
candidate, from which the energy of the photon cluster together with
the median energy density of the event times the cone area are
subtracted~\cite{Cacciari:2009dp,Aad:2010sp}, is required to be less
than 6~\GeV.  Photon candidates are rejected if they are not well
separated from the previously selected leptons, i.e. if $\Delta
R(\gamma,\ell)$ < 0.4. The overall efficiency of this photon selection
on \Zgjj EWK events is about 37\% (96\%) for photons with
$E_{\mathrm{T}}^{\gamma}>$15 (150)~\GeV.  Uncertainties in the photon selection
come from: variations in the reconstruction and identification
criteria~\cite{Aaboud:2016yuq}; changes in the isolation requirements;
and uncertainties in the photon energy scale and
resolution~\cite{Aad:2012elid}.

Jets are reconstructed from clusters of energy in the calorimeter
using the anti-$k_t$ algorithm~\cite{Cacciari:2008gp} with radius
parameter $R=0.4$. Jet energies are calibrated using energy- and
$\eta$-dependent correction factors derived using MC simulation and
validated by studying collision data~\cite{Aad:2014bia}.  Jets are
considered if they have $p_{\mathrm{T}}>30$~\GeV and $|\eta|<4.5$. To
remove jets originating from additional collisions in the same bunch
crossing, at least 50\% of the summed scalar $p_{\mathrm{T}}$ of the
tracks within a cone of size $\Delta R=0.4$ around the jet axis must
originate from the hard-interaction vertex.  This criterion is applied
only to jets with $p_{\mathrm{T}}<50$~\GeV and $|\eta|<2.4$
~\cite{PERF-2014-03}.  Jet candidates are rejected if they are not
well separated from the previously selected leptons and photons, i.e
if $\Delta R (j, \ell)<0.3$ or $\Delta R(j,\gamma)<0.4$.  Systematic
effects in jet reconstruction lead primarily to uncertainties in the
jet energy scale (JES) and resolution (JER) and are described in
Ref.~\cite{Aad:2014bia}.

The determination of the two-dimensional missing transverse-momentum
vector, $\vec{p}_{\mathrm{T}}^{\mathrm{\;miss}}$, is based on the
measurement of all topological clusters in the calorimeter and muon
tracks reconstructed in the ID and MS~\cite{Aad:2012re}. Calorimeter
cells associated with reconstructed objects, such as electrons,
photons, $\tau\rightarrow\mathrm{hadrons}+\nu$ decays, and jets, are
calibrated at their own energy scale, whereas calorimeter cells not
associated with any object are calibrated at the electromagnetic
energy scale. The magnitude of this vector is denoted by \met.
Uncertainties in the measurement of \met are derived from uncertainties 
in measurements of the contributing objects.

\subsection{Selection of \llgjj  events}
\label{sec:lep_sel}

In the charged-lepton channel, \llgjj events are required to have one photon
candidate with $E_{\mathrm{T}}^{\gamma}>15$~\GeV, a pair of
opposite-sign (OS), same-flavour leptons (electrons or muons) and at least
two reconstructed jets.

The invariant mass of the two leptons, $m_{\ell\ell}$, must be at
least 40~\GeV. The sum of the dilepton mass and the three-body
$\ell\ell\gamma$ invariant mass is required to be larger than
182~\GeV, which is approximately twice the $Z$ boson mass. This
requirement ensures that the three-body invariant mass is larger than
the $Z$ boson mass, thus suppressing the cases where the $Z$ boson
decay products radiate a photon.

The event topology of \Zgjj EWK production is characterized by the
presence of two bosons in the central region and two jets with large
rapidity difference and large dijet mass.  Different phase-space
regions are considered based on $m_{jj}$.  The inclusive region is
defined by events with no requirement on the dijet invariant mass, the
control region (CR) is defined by events with $150<m_{jj}<500$~\GeV,
and the search region (SR) is defined by requiring
$m_{jj}>500$~\GeV. The requirement of $m_{jj}>150$~\GeV suppresses the
background process of $Z\gamma+W(\to jj)$ triboson to negligible
levels. The search region definition is optimized for the best
expected significance for the \Zgjj EWK process, given amount of data.

Finally, the fiducial phase-space region optimized for sensitivity to
anomalous quartic couplings (the ``aQGC region''),
is defined by requiring events in the search region to
have a photon with $E_{\mathrm{T}}^{\gamma}>$~250~\GeV.
The expected numbers of \Zgjj EWK events in the search and aQGC regions are
22.8 $\pm$ 1.5 and 0.41 $\pm$ 0.04, respectively.

A centrality observable $\zeta$ is defined to quantify the relative
position in pseudorapidity of a particle or system of particles with respect to the
two leading jets ($j_1$ and $j_2$):

\begin{equation}
\label{eqn:centrality}
 \zeta  \equiv \left| \frac{\eta - \bar{\eta}_{jj}}{\Delta \eta_{jj}} \right| \ \ \text{with} \ \ \  \bar{\eta}_{jj}=\frac{\eta_{j_1} + \eta_{j_2}}{2}, \ \ \ \Delta \eta_{jj}= \eta_{j_1}-\eta_{j_2},
\end{equation}

where $\eta$ is the pseudorapidity of the physics object.
The centrality of the $Z\gamma$ system, \cZg, allows
discrimination between \Zgjj EWK and QCD production, with
the former contributing more at low values of \cZg. However, to
maximize the statistical power of the sample, no explicit \cZg
requirement is implemented, but rather the full \cZg distribution is
used to extract the \Zgjj cross-sections, as detailed in
Section~\ref{sec:xsec}.

\subsection{Selection of $\nu\bar{\nu}\gamma jj$ events}
\label{sec:nu_sel}

In the neutrino channel analysis, the $Z$ boson signature is
high missing transverse momemtum from the undetected
neutrino pair. Therefore, the $\nu\bar{\nu}\gamma jj$ candidate events
are required to have $\met>100$~\GeV, which corresponds to a relative \Zgjj EWK
efficiency of 85\%, along with the presence of a candidate photon with
$E_{\mathrm{T}}^{\gamma}>$150~\GeV and at least two jets. 

A lepton veto requirement (on the presence of electrons or muons as defined above)  is
applied to reduce the large contribution from $W(\ell \nu)\gamma+$jets
events. This requirement is almost 100\% efficient for \Zgjj events.

Requirements on event topology are introduced to suppress the large
background from $\gamma +$jets (where $\vec{p}_{\mathrm{T}}^{\mathrm{\;miss}}$ is usually collinear
with jets) and $W(e\nu)+$jets events. This is achieved by applying a
set of angular selection criteria:
the azimuthal difference between $\vec{p}_{\mathrm{T}}^{\mathrm{\;miss}}$ and the total transverse
momentum of the photon and the two jets should be larger than
$3\pi/4$, ($\Delta\phi(\vec{p}_{\mathrm{T}}^{\mathrm{\;miss}}, \gamma jj) > 3\pi/4$); the azimuthal
difference between $\vec{p}_{\mathrm{T}}^{\mathrm{\;miss}}$ and the photon should be larger than
$\pi/2$, ($\Delta\phi(\vec{p}_{\mathrm{T}}^{\mathrm{\;miss}}, \gamma) > \pi/2$); and the azimuthal
difference between $\vec{p}_{\mathrm{T}}^{\mathrm{\;miss}}$ and each of the two jets should be larger
than 1, ($\Delta\phi(\vec{p}_{\mathrm{T}}^{\mathrm{\;miss}}, j)>1$). Overall, these angular separation
requirements suppress the background by a factor of 40, with a
relative \Zgjj EWK efficiency of 33\%.

To enhance the \Zgjj EWK production and maximize the sensitivity to
aQGC, further event topology selections are applied: the absolute
rapidity difference between the two jets is required to be greater
than 2.5 ($|\Delta y_{jj}|>2.5$), the photon centrality must be smaller than
0.3 ($\zeta_{\gamma}<0.3$), the \pTbalance of the $\nu\bar{\nu}\gamma jj$ object, defined as

\begin{equation}
\label{eqn:ptbal}
 \pTbalance \equiv \frac{|\ptmiss+ \vec{p}_{\mathrm{T}}^{\gamma}+ \vec{p}_{\mathrm{T}}^{j_1}+ \vec{p}_{\mathrm{T}}^{j_2}|}{\met + |\vec{p}_{\mathrm{T}}^{\gamma}|+ |\vec{p}_{\mathrm{T}}^{j_1}| + |\vec{p}_{\mathrm{T}}^{j_2}|}, 
\end{equation}

must be smaller than 0.1, and the dijet invariant
mass must be greater than 600~\GeV. These event
topology requirements further reduce the background by a factor of 80,
with a relative \Zgjj EWK efficiency of 20\%.
The expected number of \Zgjj EWK events after all the selection
requirements is 0.65 $\pm$ 0.05.

\section{Background estimate and event yields}
\label{sec:background}

\subsection{Backgrounds in the charged-lepton channels}
\label{sec:background_Zllg}

The main background to the \Zgjj production processes comes from the
misidentification of hadronic jets as photons (jets faking photons) in
$Z+$jets events. This background is not well modelled by the MC
simulation. It is estimated with data using the same two-dimensional
sideband method~\cite{Aad:2010sp} used in the inclusive $Z\gamma$
cross-section measurement~\cite{bib:zg2012}. The method is based on
control regions populated by events satisfying all selection criteria but
with the candidate photon failing to satisfy some of the identification criteria
and/or the isolation requirement.

Due to the very limited number of events in the search and control
regions, the background contribution from $Z$+jets events is estimated
in an enlarged phase-space region, relaxing the dijet mass requirement
to $m_{jj}>100$~\GeV. This is the most stringent requirement on
$m_{jj}$ where the uncertainty on the background estimated is still
dominated by systematic errors. The extrapolation of the background
estimate to the search and control regions relies on the observation
that the shape of the $m_{jj}$ distribution of $Z+$jets background
events (i.e. with one jet faking a photon) in both the \powheg and \alpgen
MC samples is similar to the $m_{jj}$ distribution of $Z\gamma$ events
in \sherpa MC samples, for $m_{jj}>100$~\GeV. Therefore, the ratio of
$Z+$jets to $Z\gamma$ contribution can be considered the same in the
enlarged phase-space region as in the search and control regions.

In the enlarged phase-space region ($m_{jj}>100$~\GeV), the contribution
from $Z$+jets events is estimated with data to be ($23 \pm 6$)\% of
$Z\gamma$ events. The uncertainty is dominated by the systematic
uncertainty due to the correlation  between photon
identification and isolation requirements. This correlation is
calculated from MC simulation and the large systematic uncertainty reflects the
different responses from \sherpa, \pythia and \alpgen modelling. Other
systematic uncertainties related to control region definition, signal
contamination in control regions, and $m_{jj}$ shape difference
between $Z+$jets and $Z\gamma$ are found to be negligible compared to
the normalization uncertainty and are neglected.

Besides the $Z+$jets process, other background contributions are from
$WZ+$jets events, with a misidentification of an electron as a photon,
and $t\bar{t}\gamma$ events, with the photon emitted from
initial-state partons or final-state leptons. The yields of these two
processes are estimated from MC simulation with an uncertainty determined
by the measured cross-sections uncertainty.

\subsection{Backgrounds in the neutrino channel}
\label{sec:background_Znng}

For the neutrino channel, background events mainly arise from processes
having final states similar to the signal, from events with jets or
electrons misidentified as photons, and from events with high fake
\met (i.e. due to mismeasurement of hadronic energy deposits rather
than the presence of neutrinos in the events). The main background
processes are $W(\ell \nu)\gamma$+jets, $Z(\nu\bar{\nu})$+jets,
$\gamma+$jets and $W(e\nu)+$jets accounting for approximately 59\%,
15\%, 7\%, and 5\% of the total background, respectively.

The dominant background is $W(\ell \nu)\gamma+$jets production, where
the lepton is either not reconstructed or not identified, making the lepton
veto requirement ineffective. In particular, $W(\tau\nu)\gamma+$jets
events, with $\tau\rightarrow\mathrm{hadrons}+\nu$ decays, provide a considerable
contribution to this background.  The $W\gamma+$jets background, which
includes both the QCD and EWK components, is estimated using the
\sherpa MC samples. The normalization is determined with data. The MC
yield of $W(\ell \nu)\gamma+$jets events is corrected by constructing
a data sample from events passing the $\nu\bar{\nu}\gamma jj$
inclusive selection and requiring exactly one charged lepton in the
event (instead of vetoing them). The fraction of $W\gamma$ events in
this sample is about 80\%, and these data events (after subtracting
non-$W\gamma$ contributions from MC estimates) are used to determine a
correction factor for the MC yield of the $W\gamma$+jets sample, which is found to be 1.06. 
The difference between the background estimates extracted from \sherpa
and \alpgen MC samples is the dominant systematic uncertainty in the
$W\gamma+$jets background prediction, corresponding to a relative
uncertainty of 41\%.

The second largest source of background comes from
$Z(\nu\bar{\nu})$+jets, where a jet is misidentified as an
energetic photon. The contribution of this background is estimated
with the same two-dimensional sideband method used to determine the $Z+$jets
contribution in the charged-lepton channel. In this case, however, the
background estimate is performed directly in the phase-space region of
interest. The statistical uncertainty of 50\% is significantly larger than 
the systematic uncertainty, which amounts to 20\%.

Another important source of background is the production of
$\gamma$+jets events with fake \met. This background is
estimated with data, again using a two-dimensional sideband method. The control
regions are composed of events with low \met and/or with low
values of $\Delta\phi(\vec{p}_{\mathrm{T}}^{\mathrm{\;miss}}, j)$.
Due to the limited size of the data sample, the background estimate is performed
with a relaxed energy requirement on the photon
($E_{\mathrm{T}}^\gamma>$45~\GeV) and then extrapolated to the phase-space region
of interest using MC samples. The difference between the
extrapolation results obtained with \sherpa and \alpgen samples ($\sim$40\%) is the
dominant uncertainty for this background.

The sizeable production of $W(e\nu)+$jets is also a source of background
when the electron is misidentified as a high-energy photon.  To
estimate this background, first the fake rate of $e \rightarrow
\gamma$ misidentification is extracted from data using electrons from
$Z\to ee$ events.
Then the $W(e\nu)+$jets background contribution is estimated by
applying this fake rate to events passing the full event selection but
choosing a high-energy electron instead of a photon. The main
uncertainty comes from the limited size of the control sample
and equals 43\%.

The background contribution from $Z(\tau \tau)\gamma$ is also estimated with MC samples
and found to be less than 1\%.

\subsection{Expected and observed event yields}
\label{sec:EventYield}

Table~\ref{tab:ZgYield} summarizes the event yield for the
charged-lepton channel, including details from the various signal and
background processes.  Three different phase-space regions are
presented: inclusive $Z\gamma\ +\ge$ 2 jets selection, CR and SR.  A
breakdown of the sources of systematic uncertainty in the CR and SR is
given in Table~\ref{tab:systXSyield}.  Table~\ref{tab:zllgresults_vbs}
summarizes the event yield for both the charged-lepton and neutrino
channels in the aQGC region with systematic uncertainties summarized in
Table~\ref{tab:systQGC}.  In the aQGC region, relative uncertainties
in the yield in the charged-lepton channel are the same as those in
the SR except for that arising from photon selection. This component
is larger due to the higher value required for the photon transverse energy
in the aQGC region.

Figure~\ref{fig:photon_pt} displays the transverse energy of the
photon after various selection requirements (inclusive selection,
control and search regions); Figure~\ref{fig:ngoodJ} shows the numbers
of selected jets in the control and search regions;
Figure~\ref{fig:Jmjj} shows the distribution of the dijet mass for the
inclusive selection; and finally Figure~\ref{fig:templ_SR_CR} displays
the distributions of $\cZg$ in the inclusive, control and search
regions.
Corresponding kinematic
distributions for the neutrino channel are shown in
Figure~\ref{fig:ZnunuGKinPlots150vbs}.  The \sherpa MC prediction is
found to describe the data well for all these variables and in all
phase-space regions.

\begin{table}[h]
\caption{Summary of events observed in data and estimated composition for the $Z(e^+e^-)\gamma jj$ and
$Z(\mu^+\mu^-)\gamma jj$  production processes. The $Z+$jets contribution in this table  is taken as a fixed fraction, $(23 \pm 6)\%$, of $N_{\text{$Z
\gamma$ QCD}}$. The last line corresponds to the sum of the two previous lines ($N_{\text{$Z
\gamma$ QCD}}$ + $N_{\text{$Z \gamma$ EWK}} $). The uncertainties correspond to the statistical and systematic uncertainties added in quadrature.}
\begin{center}
\begin{tabular}{lr@{$\,\pm\,$}lr@{$\,\pm\,$}lr@{$\,\pm\,$}lr@{$\,\pm\,$}lr@{$\,\pm\,$}lr@{$\,\pm\,$}l}
 \hline
   & \multicolumn{4}{c}{Inclusive region} &  \multicolumn{4}{c}{Control region} & \multicolumn{4}{c}{Search region} \\
   & \multicolumn{4}{c}{$Z(\ell^+\ell^-)\gamma\ + \ge 2$ jets} & \multicolumn{4}{c}{$150<m_{jj}<500$~\GeV} & \multicolumn{4}{c}{$m_{jj}>500$~\GeV}\\  
   & \multicolumn{2}{c}{$e^+e^-\gamma jj$} & \multicolumn{2}{c}{$\mu^+\mu^-\gamma jj$} & \multicolumn{2}{c}{$e^+e^-\gamma jj$} & \multicolumn{2}{c}{$\mu^+\mu^-\gamma jj$} &
 \multicolumn{2}{c}{$e^+e^-\gamma jj$} & \multicolumn{2}{c}{$\mu^+\mu^-\gamma jj$} \\
 \hline \hline
 Data            &  \multicolumn{2}{c}{781}   &    \multicolumn{2}{c}{949}  &   \multicolumn{2}{c}{362}     & \multicolumn{2}{c}{421}   & \multicolumn{2}{c}{58}      & \multicolumn{2}{c}{72} \\
 \hline \hline
 $Z+$jets bkg.  &                    134 & 36     & 154 & 42    & 57 & 16     & 67 & 18    & 8.5 & 2.5  & 9.4 & 2.7 \\
 Other bkg. ($t\bar{t}\gamma$, $WZ$)  & 88 & 17     & 91  & 18    & 47 & 9      & 46 & 9     & 5.8 & 1.1  & 5.0 & 1.0  \\
\hline
 $N_{\text{data}} -N_{\text{bkg}} $ &    559 & 46 & 704 & 53 & 258 & 24 & 308 & 27     & 44 & 7  & 58 & 8   \\
\hline \hline
 $N_{\text{$Z \gamma$ QCD}}$ (\sherpa MC)    & 583  & 41   &    671  & 47     &  249 &  24    & 290 & 26   & 37   & 5    & 41   &  5   \\
 $N_{\text{$Z \gamma$ EWK}} $ (\sherpa MC)   & 25.4 & 1.5  &    27.3 & 1.7    &  8.6 &  0.6   & 9.3 & 0.6  & 11.2 & 0.8  & 11.6 &  0.7 \\
\hline
 $N_{\text{$Z \gamma$}} $ (\sherpa MC)       & 608  & 42   &    698  & 49     &  258 &  25    & 299 & 27   & 48   & 6    & 53   &  6     \\
\hline \hline
\end{tabular}
\end{center}
\label{tab:ZgYield}
\end{table}

\begin{table}[h]
  \caption{Summary of the dominant experimental systematic uncertainties in the event yield
   in the CR and SR,
   for the electron (muon when different) channel and for
   the signal and main background
   components.}
 \begin{center}
\begin{tabular}{lcccccc}
\hline
  Source of  & \multicolumn{2}{c}{EWK yield [\%]}    & \multicolumn{2}{c}{QCD yield [\%]}   &  \multicolumn{2}{c}{Bkg. yield [\%]}  \\
  uncertainty   & CR   & SR                     & CR   & SR                     & CR & SR  \\
\hline
  Trigger &              \multicolumn{2}{c}{}        &   \multicolumn{2}{c}{0.2 (0.4)}      &  \multicolumn{2}{c}{} \\
  Pile-up &              \multicolumn{2}{c}{}        &   \multicolumn{2}{c}{0.6}            &  \multicolumn{2}{c}{} \\
  Lepton selection &     \multicolumn{2}{c}{}        &   \multicolumn{2}{c}{3.8 (2.3)}      &  \multicolumn{2}{c}{} \\
  Photon selection &     \multicolumn{2}{c}{}        &   \multicolumn{2}{c}{1.6}            &  \multicolumn{2}{c}{} \\
  Jet reconstruction &                        1.1         & 2.5        & 5.0          & 12         & 4.9        & 12 \\
  Bkg. 2D sideband   &                        -             & -            & -              & -               & 26         & 26 \\
\hline 
  Total experimental &  4.3 (3.1) & 4.9 (3.8) & 6.5 (5.8) & 13 (12) & 27 (27) & 29 (29) \\       
  Theory                         &   5.2         & 8.7         & 5.6        & 3.8           & 5.6             & 3.8 \\
\hline
\end{tabular}
\end{center}
  \label{tab:systXSyield}
\end{table}

\begin{table}[h]
\caption{Summary of events observed in data and estimated composition
  of the background for the $Z(\ell\ell)\gamma jj$ and
  $Z(\nu\bar{\nu})\gamma jj$ production processes in the aQGC region. The last
  line corresponds to the sum of the two previous lines ($N_{\text{$Z
      \gamma$ QCD}}$ + $N_{\text{$Z \gamma$ EWK}} $). The quoted
  uncertainty corresponds to the total statistical plus systematic
  uncertainty added in quadrature. }
\begin{center}
\begin{tabular}{lcc}
\hline
  & \multicolumn{2}{c}{aQGC region}\\
  &  $m_{jj}>500$~\GeV &  $m_{jj}>600$~\GeV \\
  & $E_{\mathrm{T}}^{\gamma} > 250$~\GeV  &  $E_{\mathrm{T}}^{\gamma} > 150$~\GeV \\
  & $\ell^+\ell^-\gamma jj$ & $\nu\bar{\nu}\gamma jj$ \\
\hline \hline
  Data & $2$  &  4 \\
\hline \hline 
  $Z$+jets background & $0.28\pm0.08$  & $0.3\pm0.2$ \\
  $W(\ell \nu)\gamma+$jets background & - & $1.1 \pm 0.5 $\\
  $\gamma$+jets background  & - & $0.13\pm0.08$\\
  $W(e\nu)+$jets background  & - & $0.09\pm 0.04 $\\
  $t\bar{t}\gamma$, $WZ$  background  & $0.02\pm0.01 $ & -\\
\hline
  $N_{\text{data}} -N_{\text{bkg}} $ & $1.7\pm 1.4$ & $2.4\pm2.0$ \\
\hline \hline
  $N_{\text{$Z \gamma$ QCD}}$ (\sherpa MC) & $1.2\pm 0.4 $ & $0.29\pm0.07$\\
  $N_{\text{$Z \gamma$ EWK}}$ (\sherpa MC) & $0.41\pm0.04$ & $0.65\pm0.05$\\
\hline
  $N_{\text{$Z \gamma$}}$ (\sherpa MC)  & $1.6\pm0.4$ & $0.9\pm0.1$\\
\hline \hline
\end{tabular}
 \label{tab:zllgresults_vbs}
\end{center}
\end{table}

\begin{table}[h]
 \caption{Summary of the main relative uncertainties in the MC-based
   EWK and QCD yields for the electron (muon when different) and
   neutrino channels in the aQGC region. The uncertainties in
   the $Z$+jets, $W(\ell \nu)\gamma+$jets, $\gamma$+jets, and
   $W(e\nu)+$jets yields, estimated with data, are detailed in
   the text.}
\begin{small}
\begin{center}
\begin{tabular}{lcccc}
\hline
  Source of      & \multicolumn{2}{c}{EWK yield [\%]} &  \multicolumn{2}{c}{QCD yield [\%]}  \\
  uncertainty  & $\ell^+\ell^-$ channel & $\nu\bar{\nu}$ channel
    &  $\ell^+\ell^-$ channel & $\nu\bar{\nu}$ channel \\
\hline
  Trigger &                                   0.2 (0.4) & 2 &  0.2 (0.4) & 2  \\
  Pile-up &                                    \multicolumn{4}{c}{0.6}\\
  Lepton selection &                          3.8 (2.3) & -  & 3.8 (2.3) & -\\
  \met reconstruction &                           -  & 0.4  & - & 0.4 \\
  Photon selection &                          6.5 & 3.3  & 6.5 & 3.3 \\
  Jet reconstruction &                         2.5 &  3.2      &  12 & 3.2 \\
\hline 
  Total experimental &                        8.0 (7.4)  & 5.1  & 13  & 5.1 \\ 
 Theory &                                                8.7    &  4.1    & 3.8   & 4.1 \\
\hline
\end{tabular}
\end{center}
 \label{tab:systQGC}
\end{small}
\end{table}

\begin{figure}[h]
  \begin{center}
    \includegraphics[width=0.49\columnwidth]{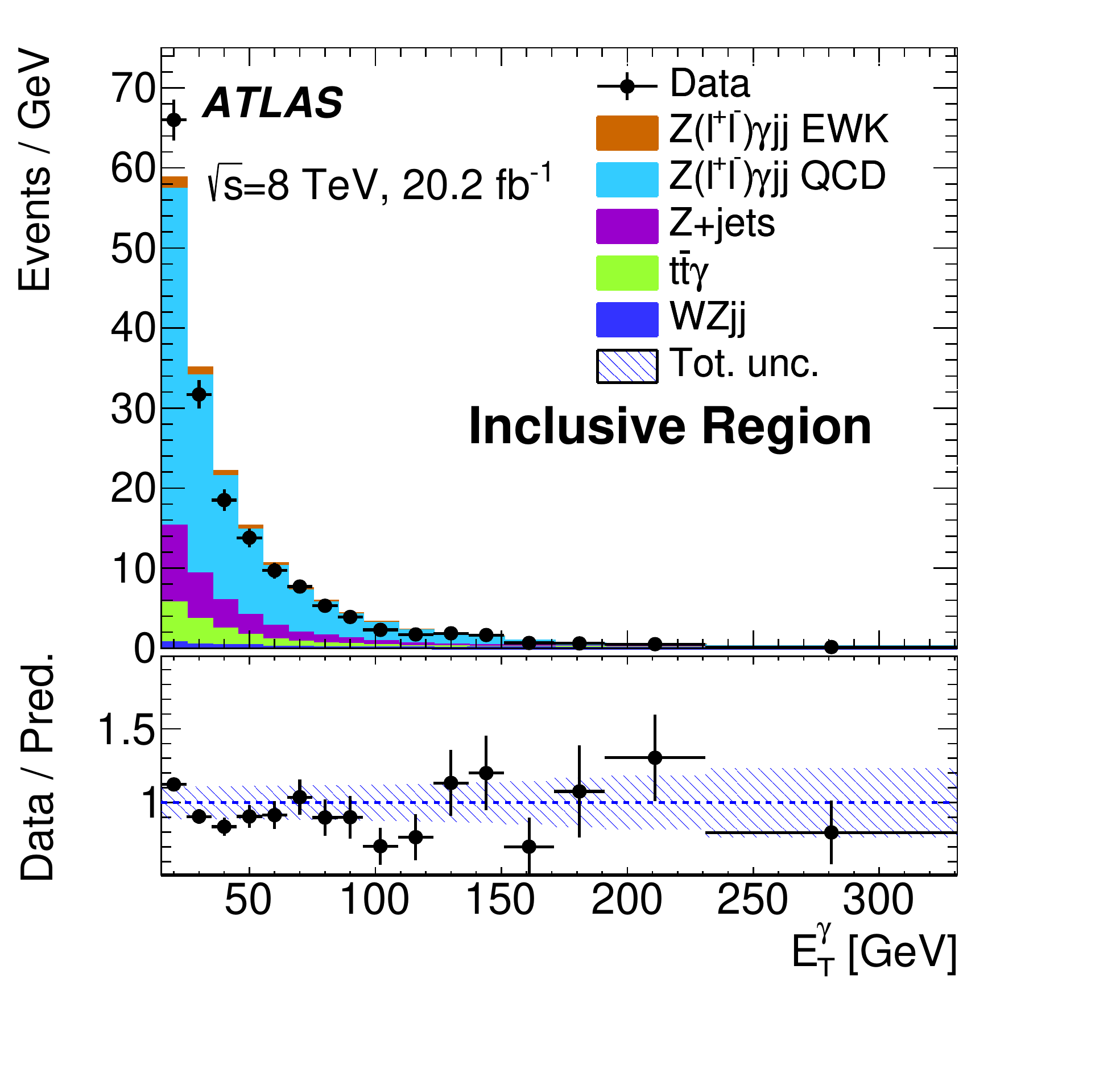}
    \includegraphics[width=0.49\columnwidth]{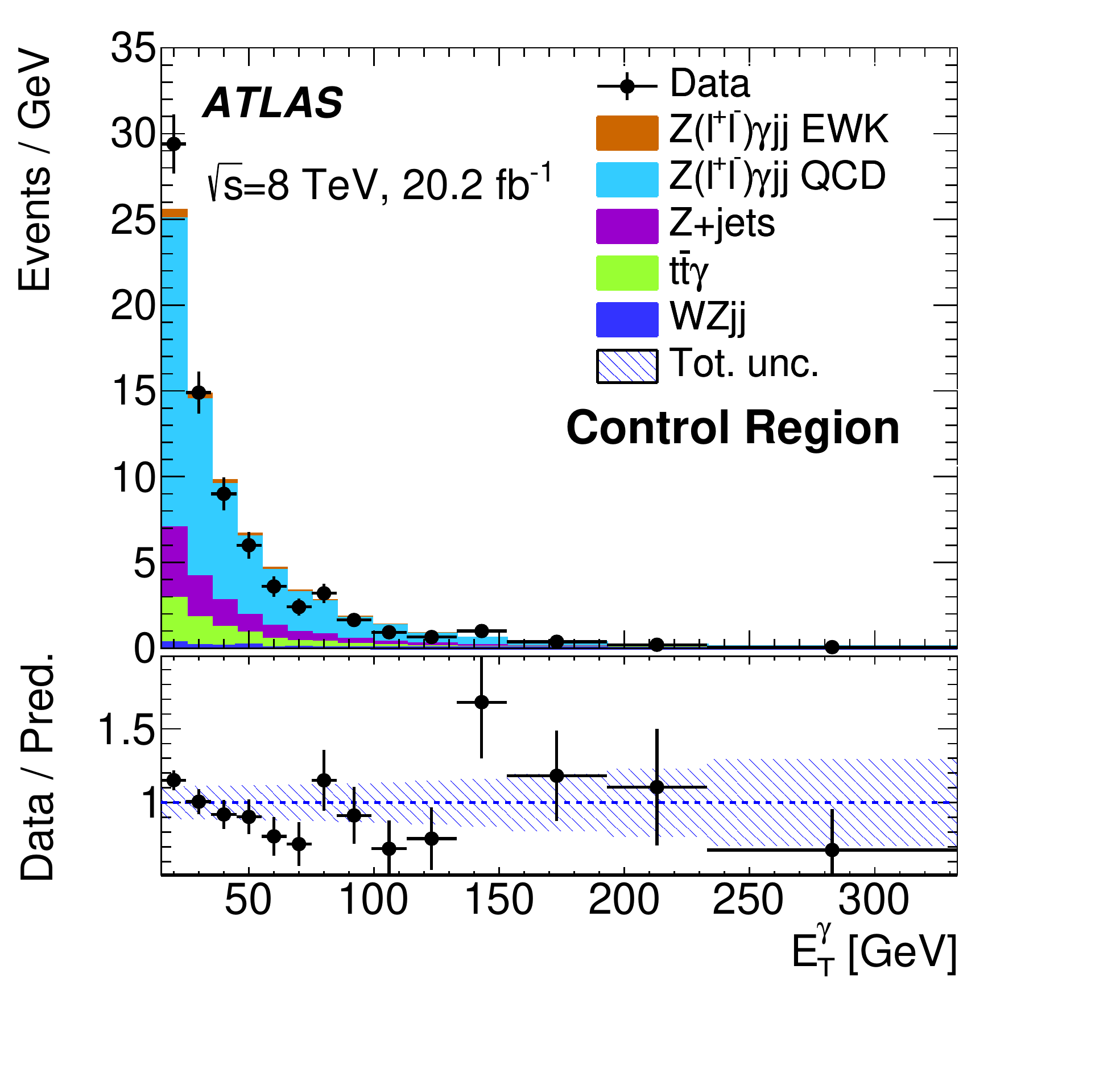}\\
    \includegraphics[width=0.49\columnwidth]{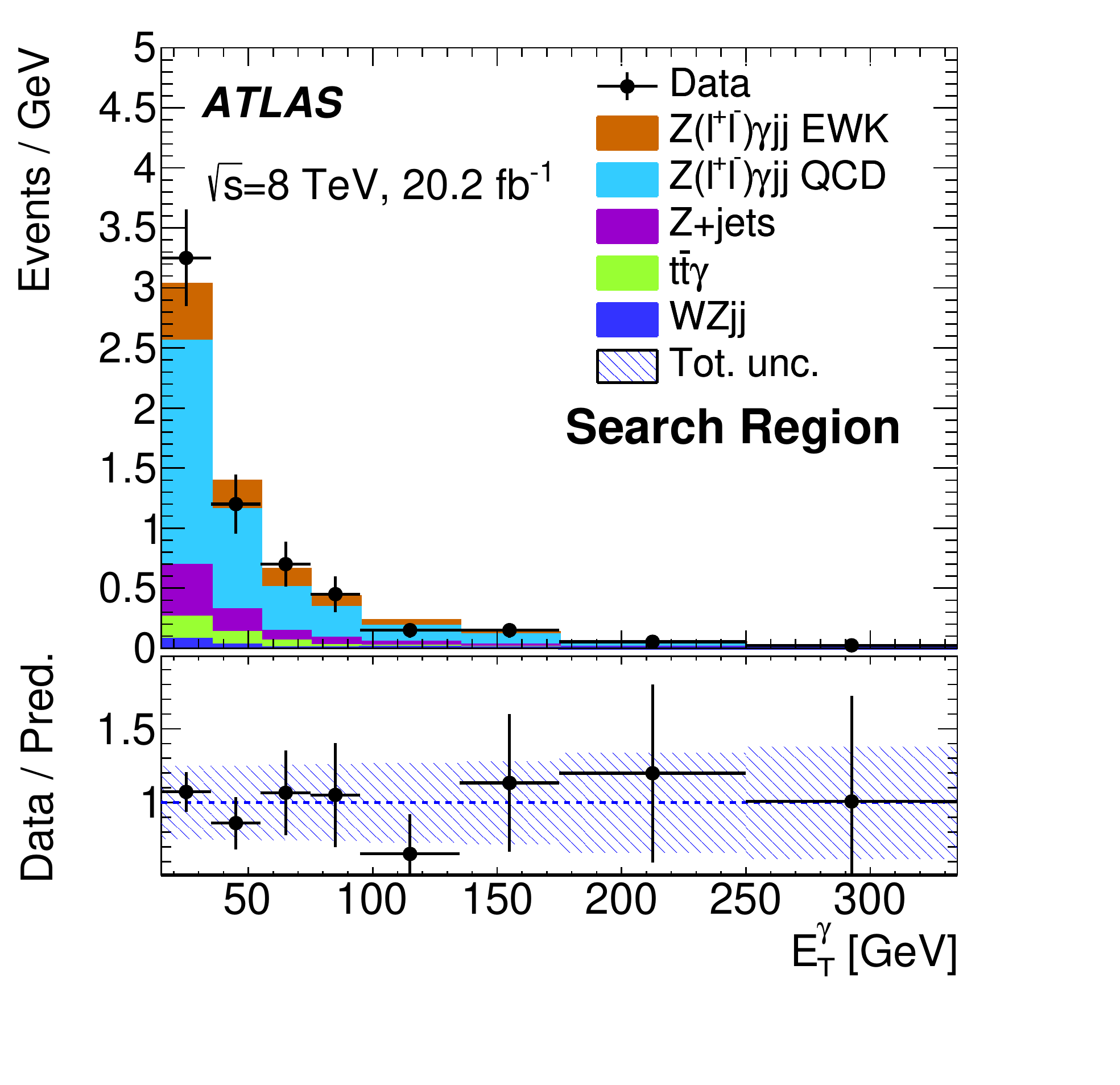}
    \caption{Distributions of the photon transverse energy for the electron and muon channels
      added together in the inclusive region (top left), in the
      control region (top right) and in the search region (bottom),
      for the data (black points), and for the signal process and
      various background components (coloured templates) before any
      fit is done. 
      The ratio of the  data to the sum of all pre-fit expected contributions (``Pred.'') 
      is shown below  each histogram. The hatched blue band shows the
      systematic and statistical uncertainty added in quadrature (``Tot. unc.'') in the signal and background
      prediction, 
      while the error bars on the data points represent the statistical uncertainty of the data set. 
      The number of events in each bin is divided by the bin width. The last bin also
      includes events beyond the range shown. } 
    \label{fig:photon_pt}
  \end{center}
\end{figure}

\begin{figure}[h]
  \begin{center}
    \includegraphics[width=0.49\columnwidth]{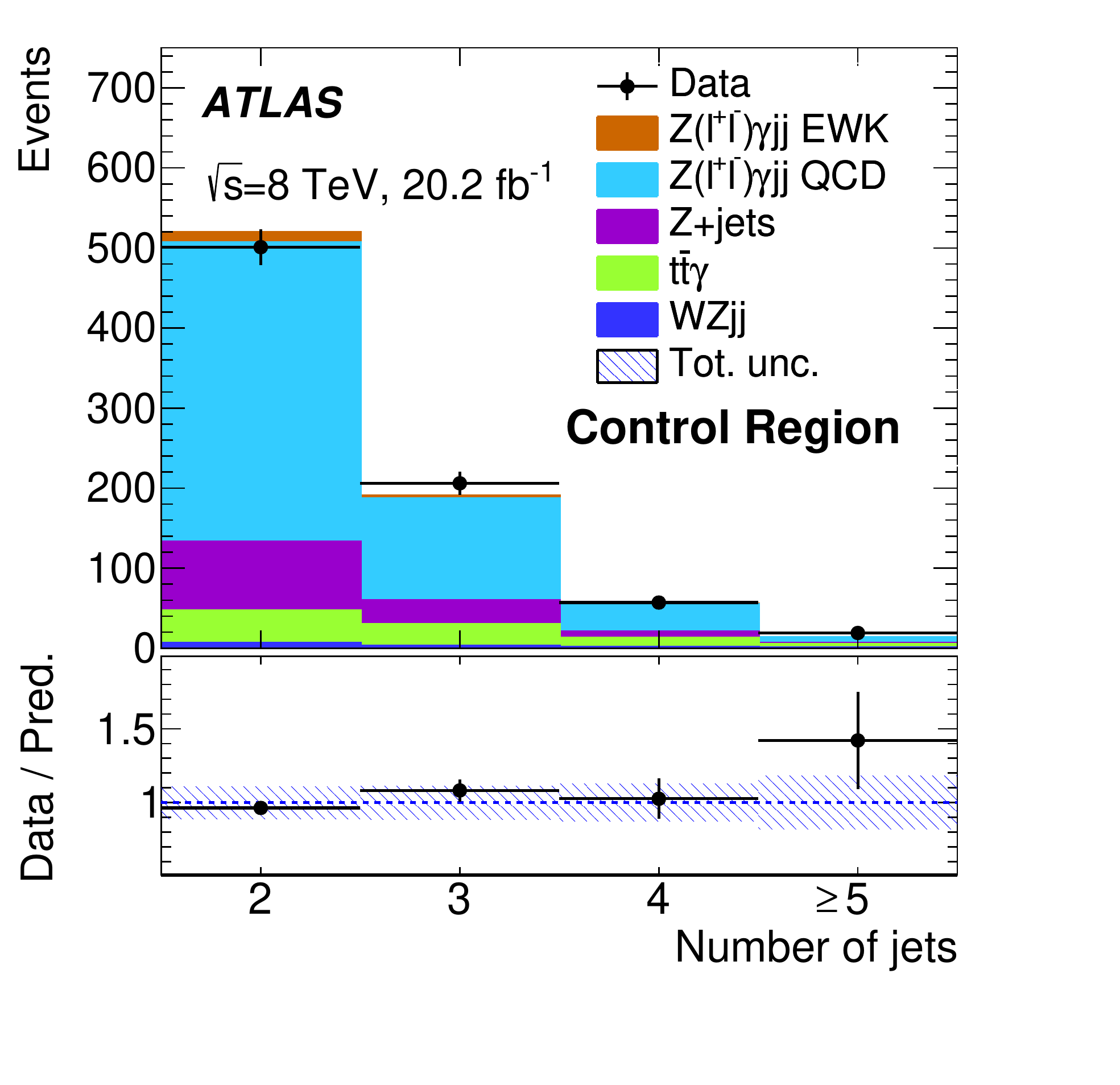}
    \includegraphics[width=0.49\columnwidth]{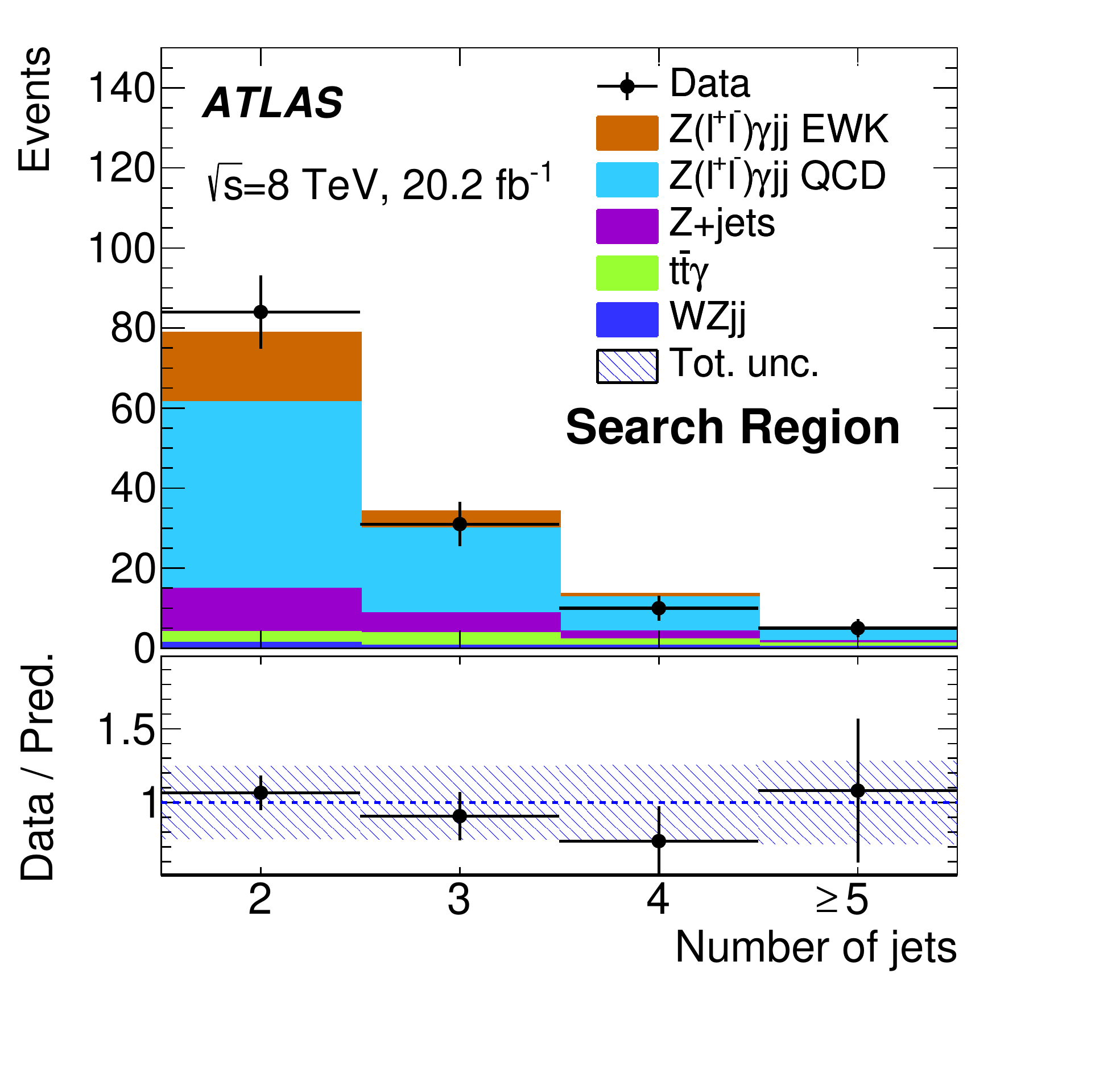}
    \caption{ Distributions of the number of jets passing the selection for the electron and
      muon channels added together in the
      control region (left) and in the search region (right),
      for the data (black points), and for the signal process and
      various background components (coloured templates) before any
      fit is done. 
      The ratio of the data to the sum of all pre-fit expected contributions (``Pred.'') 
      is shown below  each histogram. The hatched blue band shows the
      systematic and statistical uncertainty added in quadrature (``Tot. unc.'') in the signal and background
      prediction, 
      while the error bars on the data points represent the statistical uncertainty of the data set. 
      The last bin also includes events beyond the range shown. } 
    \label{fig:ngoodJ}
  \end{center}
\end{figure}

\begin{figure}[!htb]
  \begin{center}
    \includegraphics[width=0.80\columnwidth]{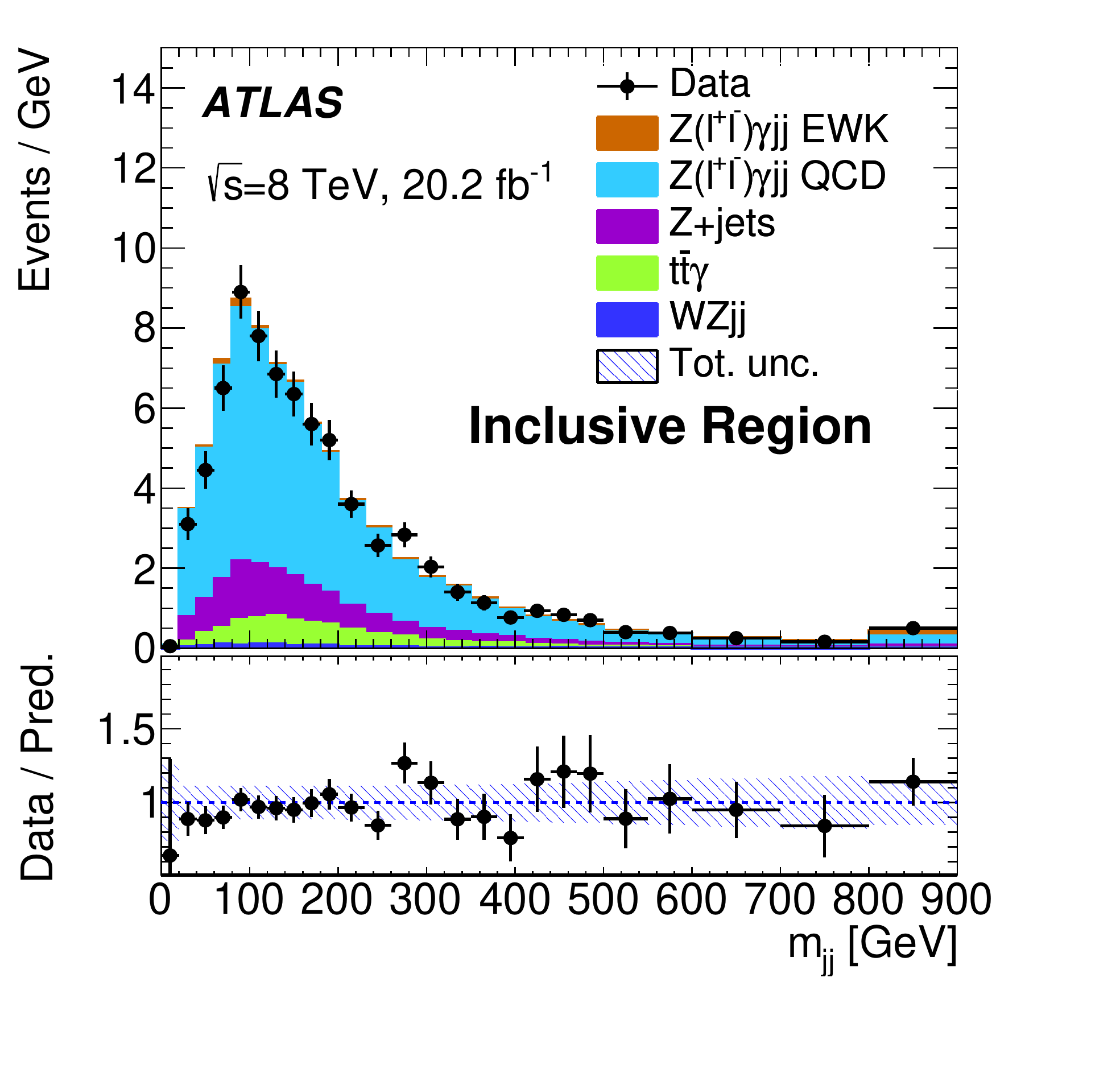}
    \caption{Distributions of the dijet invariant mass of the two
      leading jets for the electron and muon channels
      added together in the inclusive region,
      for the data (black points), and for the signal process and
      various background components (coloured templates) before any
      fit is done. 
      The ratio of the data to the sum of all pre-fit expected contributions (``Pred.'')
      is shown below the histogram. The hatched blue band shows the
      systematic and statistical uncertainty added in quadrature (``Tot. unc.'') in the signal and background
      prediction, 
      while the error bars on the data points represent the statistical uncertainty of the data set. 
      The number of events in each bin is divided by the bin width. The last bin also
      includes events beyond the range shown. } 
    \label{fig:Jmjj}
  \end{center}
\end{figure}

\begin{figure}[!htb]
  \begin{center}
    \includegraphics[width=0.49\columnwidth]{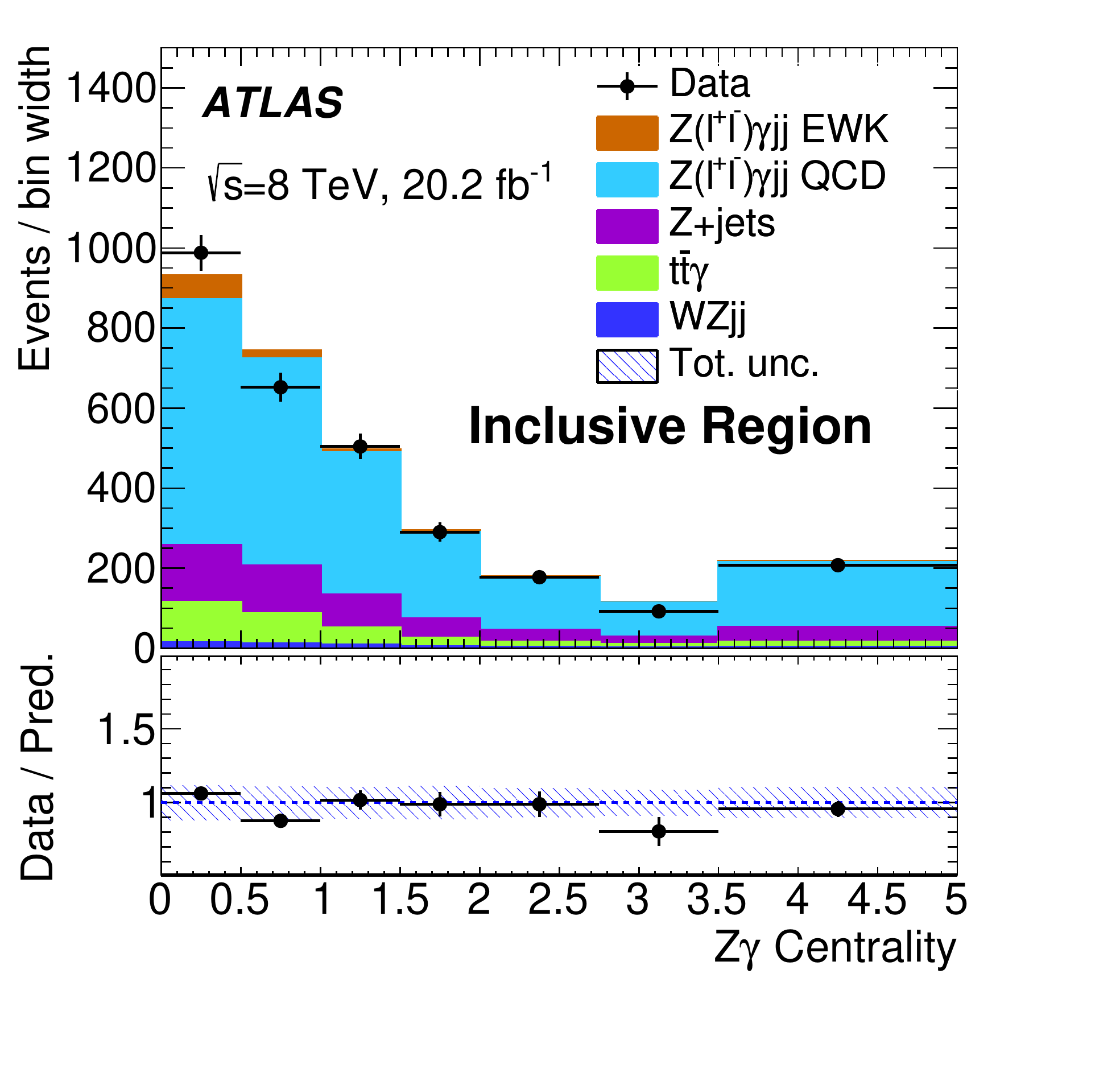}
    \includegraphics[width=0.49\columnwidth]{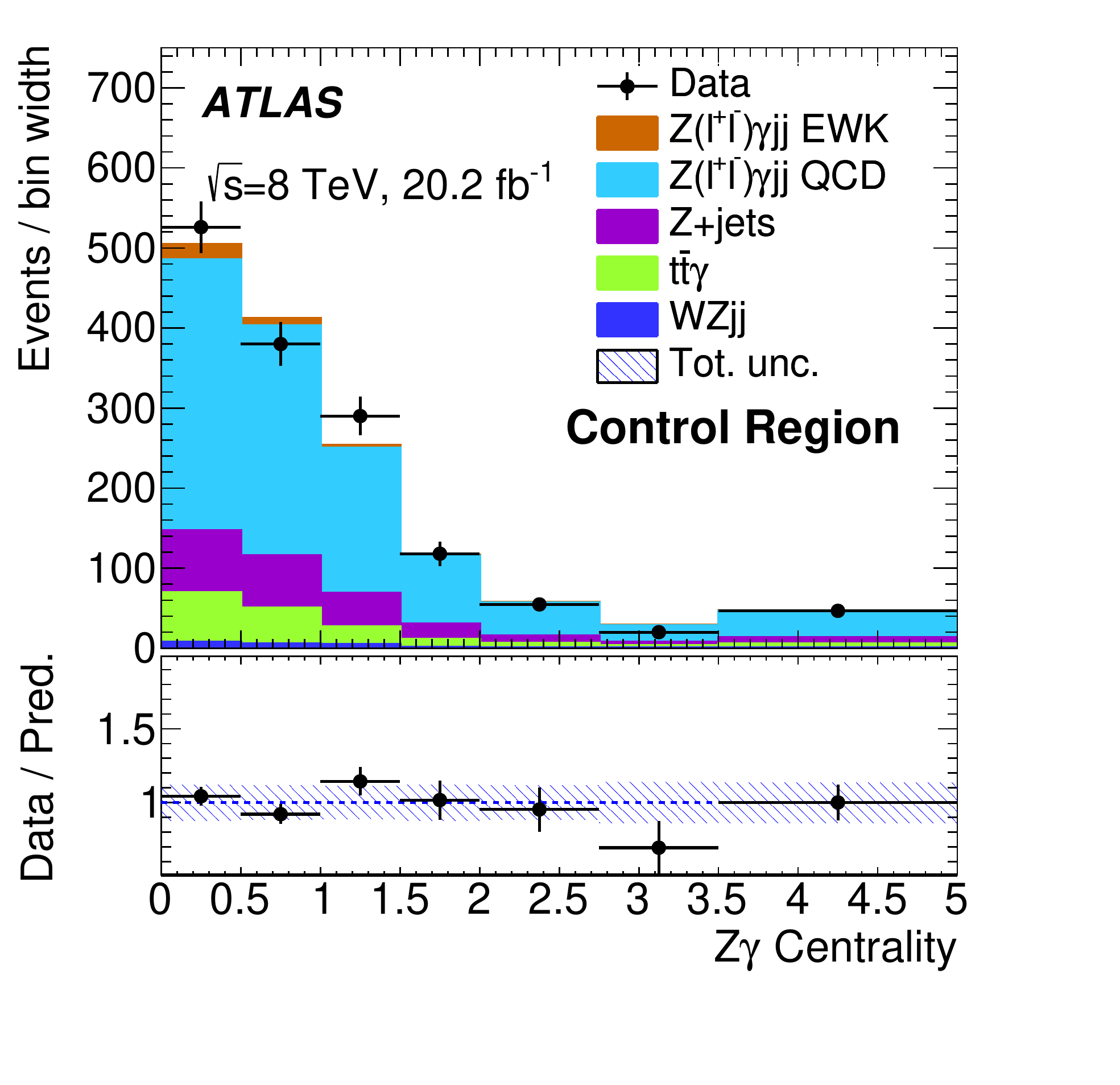}\\
    \includegraphics[width=0.49\columnwidth]{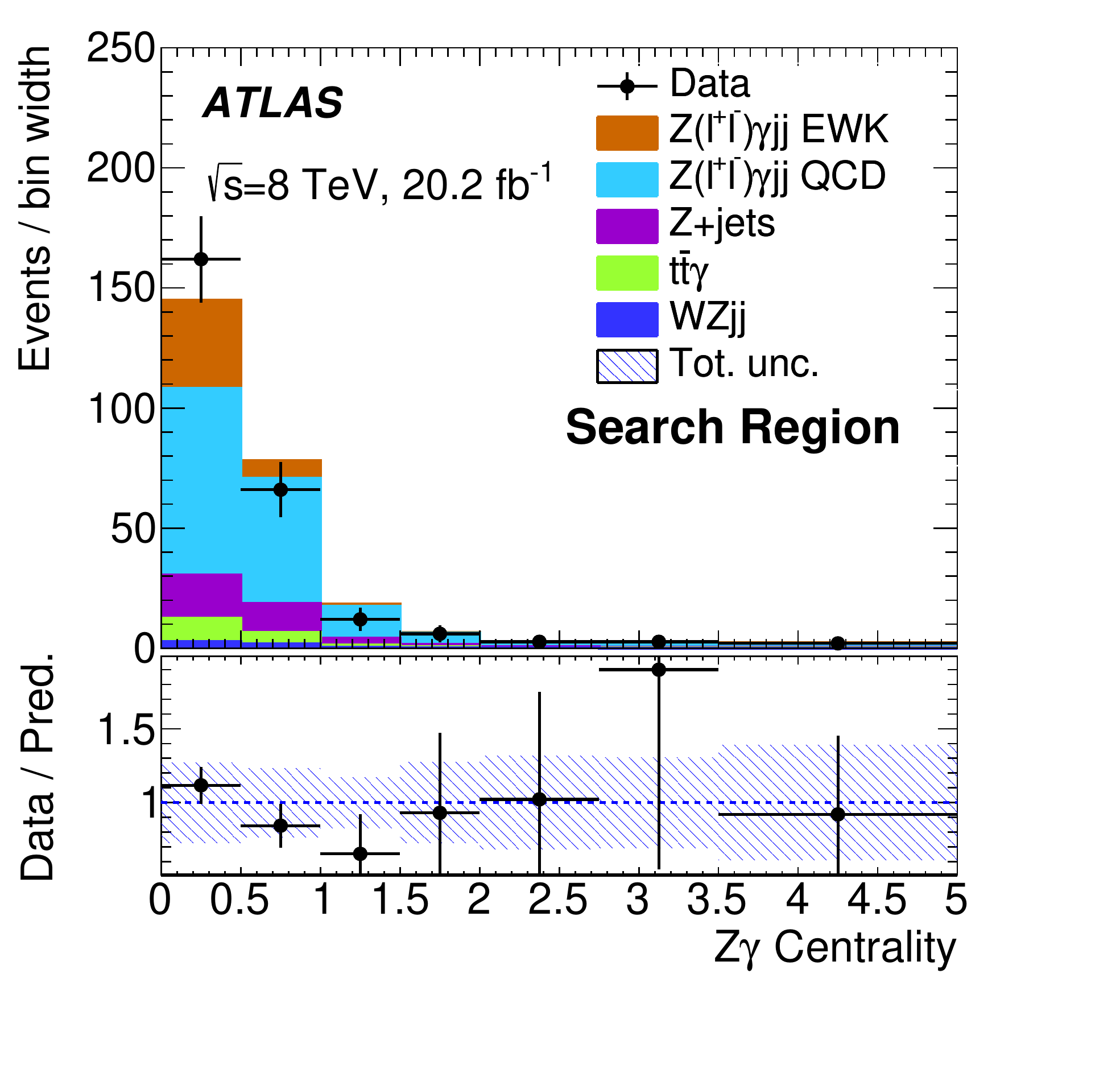}
    \caption{Distributions of centrality of the $Z\gamma$ system, $\cZg$, for the electron and muon channels
      added together in the inclusive region (top left), in the
      control region (top right) and in the search region (bottom),
      for the data (black points), and for the signal process and
      various background components (coloured templates) before any
      fit is done. 
      The ratio of the data to the sum of all pre-fit expected contribution (``Pred.'')
      is shown below each histogram. The hatched blue band shows the
      systematic and statistical uncertainty added in quadrature
      (``Tot. unc.'') in the signal and background prediction, while
      the error bars on the data points represent the statistical
      uncertainty of the data set.  The number of events in each bin
      is divided by the bin width. The last bin also includes events
      beyond the range shown. }
    \label{fig:templ_SR_CR}
  \end{center}
\end{figure}

\begin{figure}[!htb]
	\begin{center}
		\includegraphics[width=0.49\textwidth]{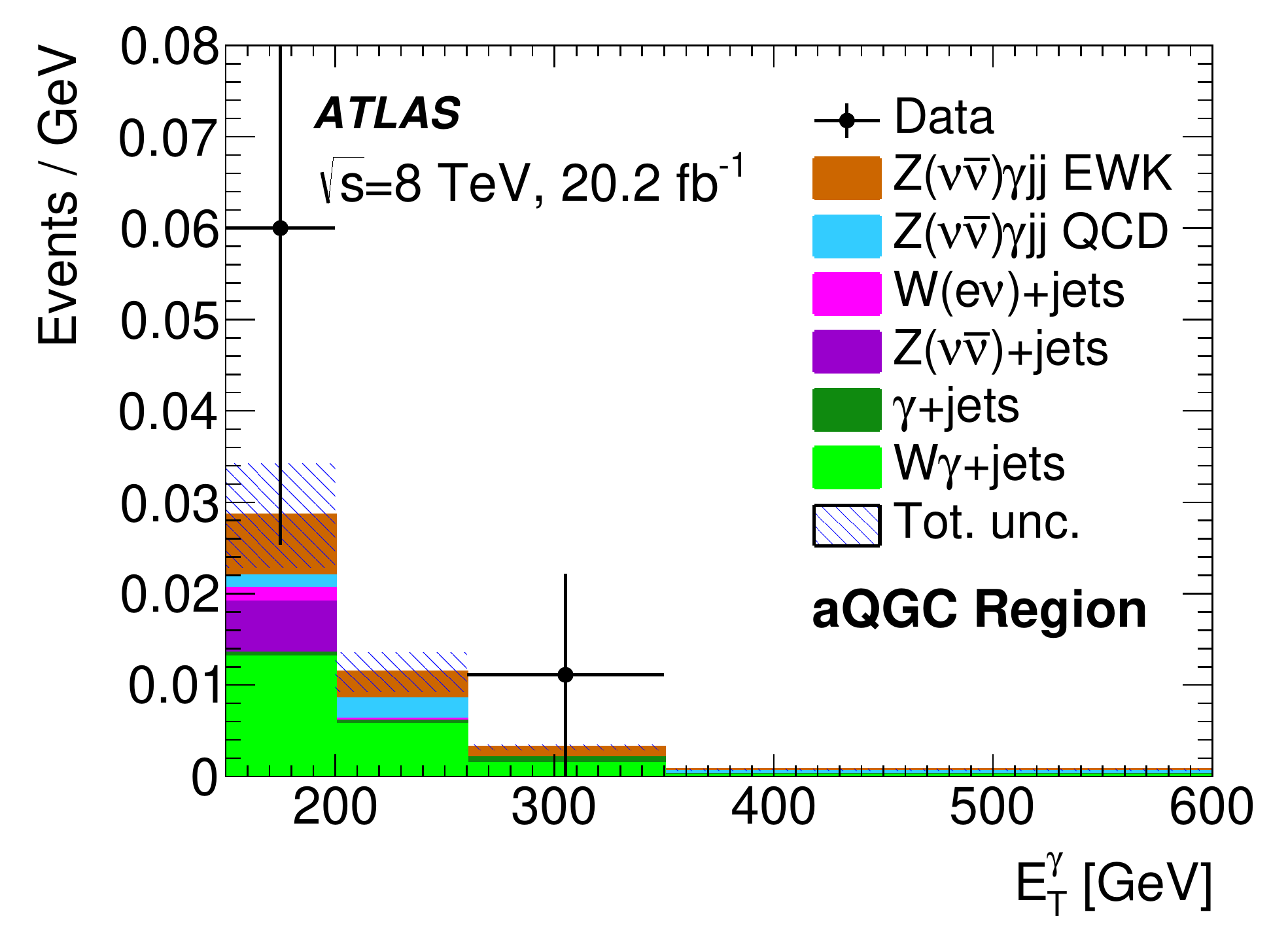}
		\includegraphics[width=0.49\textwidth]{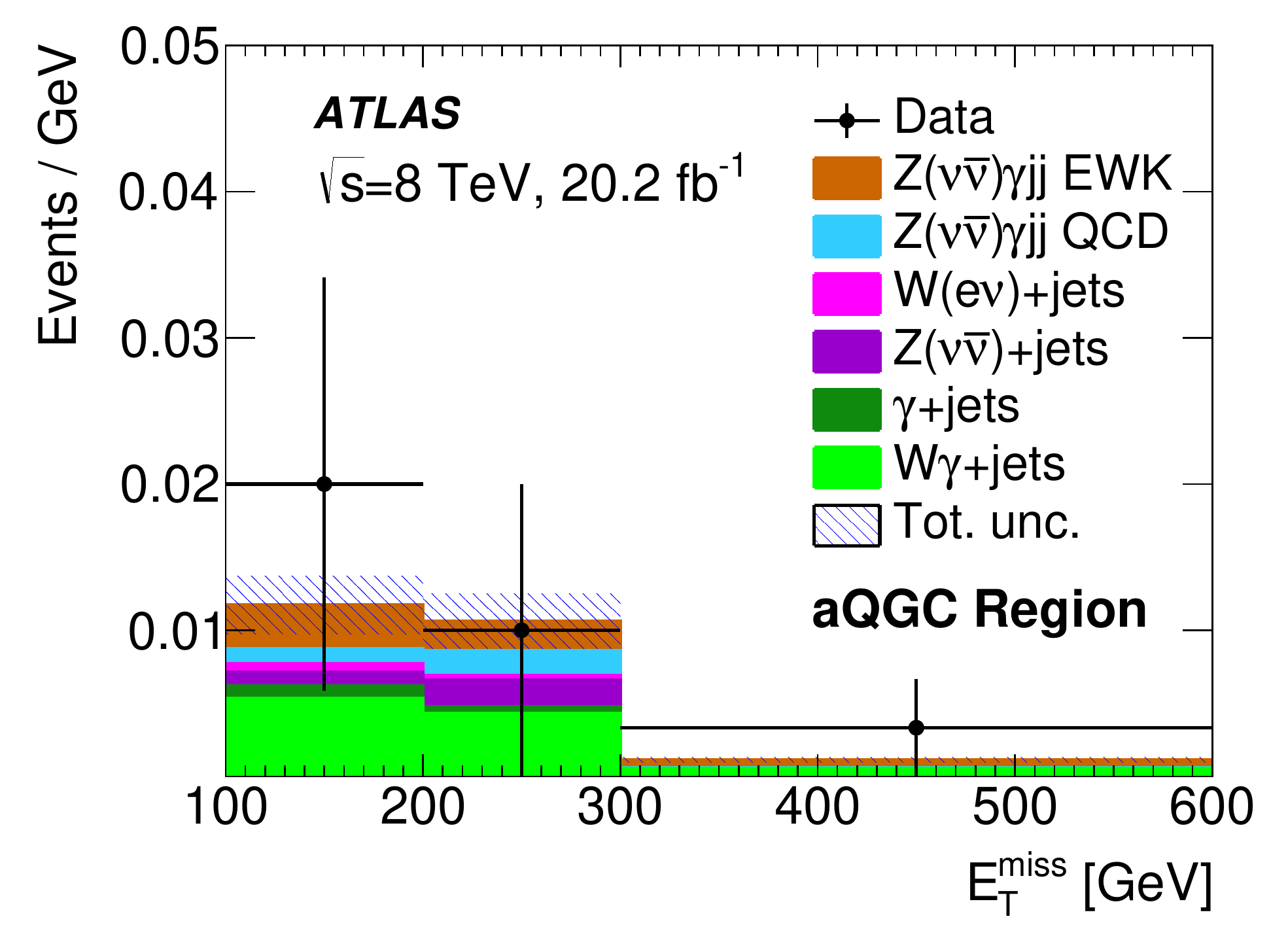}
		
	\end{center}
	\caption{Distributions of the photon transverse energy $E_{\mathrm{T}}^{\gamma}$ (left)
          and $E_{\mathrm{T}}^{\mathrm{miss}}$ (right) for the neutrino channel in the aQGC
          region with $E_{\mathrm{T}}^{\gamma} \geq 150$~\GeV
          for the data (black points), and for the signal process and
          various background components (coloured templates).
          The hatched blue band shows the systematic and statistical
          uncertainty added in quadrature (``Tot. unc.'') in the
          signal and background prediction, while the error bars on
          the data points represent the statistical uncertainty of the
          data set.  The number of events in each bin is divided by
          the bin width. The last bin also includes events beyond the
          range shown. }
	\label{fig:ZnunuGKinPlots150vbs}
\end{figure}

\newpage

\FloatBarrier
\section{Fiducial $Z\gamma jj$ cross-section measurements in the charged-lepton channel}
\label{sec:xsec}

In this section, the extraction of the \Zgjj EWK production cross-section
in the SR of the charged-lepton channel along with
the \Zgjj total (EWK+QCD) production cross-section in both the search
and control regions of the charged-lepton channel is reported. Given the very
limited number of signal events expected after the $\nu\bar{\nu}\gamma jj$
selection, that topology is only used in the search for
anomalous quartic couplings described in Section~\ref{sec:aqgc}.

Fiducial regions are defined at the particle level, using stable
particles --- those with proper lifetime longer than 10 ps --- before
their interaction with the detector. Prompt lepton four-momenta --- not
from hadron or $\tau$ decays --- are obtained through a four-vector
sum of leptons with radiated photons within a cone of radius $\Delta R$
= 0.1 around the leptons (``dressed leptons'').  Jets are
reconstructed with the anti-$k_t$ jet reconstruction algorithm with
radius parameter $R$ = 0.4 using stable particles, excluding muons and
neutrinos. The photon isolation energy is taken as the energy of the
jet matching the photon ($\Delta R(j,\gamma)<0.3$), with the photon
energy subtracted. These fiducial phase-space regions where the
measurements are performed are defined to be as close as possible to
the experimental phase-space regions, corresponding to the
reconstructed-event selection described in
Section~\ref{sec:selection}.  This minimizes the extrapolation to the
particle-level phase space by the MC simulation. 
Table~\ref{tab:selection_lep} summarizes the selection that are applied to obtain the various fiducial regions.
The fraction of events in the SR fiducial region passing the reconstructed event selection is
about 94\% for the charged-lepton channel.

The parton-level selection, used to calculate the
predicted cross-section with the \vbfnlo{} MC event generator, is basically identical to
the particle-level selection described above, with jet selection
requirements directly applied to the outgoing partons.

In order to compare the measured fiducial cross-section (at particle
level) with the NLO theory predictions by \vbfnlo{} (at parton level)~\cite{bib:ZyNLO},
a correction factor \apart is derived. Such a correction accounts for
differences between the parton- and particle-level phase-space
regions. It is defined as the ratio of the number of generated events
in the parton-level phase-space region to the number of events in the
particle-level phase-space region, and it depends on fragmentation and
hadronization models implemented in MC event generators. Using the \Zgjj EWK
\sherpa MC simulation, \apart is found to be 1.02 and 0.86 for the CR and
SR respectively (with negligible statistical uncertainty).

\begin{table}[h!]
  \centering
  \caption{Charged-lepton channel phase-space region
  definitions at particle level (parton level when different) for both
  $pp\to Z\gamma jj$ EWK and QCD production. If there are more than
  two jets/final-state partons, the two highest transverse momentum ones are
  considered.}
  \begin{tabular}{lc}
    \hline
Objects &  Particle- (Parton-) level selection \\
\hline 
Leptons & $p_{\mathrm{T}}^{\ell} > 25$~\GeV and $|\eta^\ell|<2.5$ \\
 & Dressed leptons, OS charge \\
\hline
Photon (kinematics) & $E_{\mathrm{T}}^{\gamma}>15$~\GeV, $|\eta^{\gamma}|<2.37$ \\
                 & $\Delta R(\ell, \gamma)>0.4$ \\
\hline
Photon (isolation) & $E_{\mathrm{T}}^{\mathrm{iso}} < 0.5 \cdot E_{\mathrm{T}}^{\gamma}$ (no isolation) \\
\hline
FSR cut & $m_{\ell\ell}+m_{\ell\ell \gamma}>182$~\GeV  \\
        & $m_{\ell\ell}>40$ \GeV \\
\hline
Particle jets (Outgoing partons) & At least two jets (outgoing partons) \\
($j$ = jets) & $E_{\mathrm{T}}^{j(p)}>30$ \GeV, $|\eta^{j(p)}|<4.5$ \\
($p$ = outgoing quarks or gluons)  & $\Delta R(\ell, j(p))>0.3$ \\
                                    & $\Delta R(\gamma, j(p))>0.4$ \\
\hline
Control region (CR) &  $150<m_{jj(pp)}<500$ \GeV \\
\hline
Search region (SR) &  $m_{jj(pp)}>500$ \GeV \\
\hline
 aQGC region &  $m_{jj(pp)}>500$ \GeV \\
  &  $E_{\mathrm{T}}^{\gamma}>250$ \GeV \\
\hline
  \end{tabular}   \label{tab:selection_lep}
\end{table}

\subsection{Fiducial electroweak production cross-section determination}
\label{sec:XSewk}
The determination of the fiducial cross-section is carried out using the signal strength parameter $\mu$
\begin{equation}
\label{eq:mu}
\mu= \frac{N_{\mathrm{data}}^{\mathrm{signal}}}{N_{\mathrm{MC}}^{\mathrm{signal}}} = \frac{\sigma_{\mathrm{data}}}{\sigma_{\mathrm{MC}}},
\end{equation}
where $N_{\mathrm{data}}^{\mathrm{signal}}$ is the signal yield in the data
and $N_{\mathrm{MC}}^{\mathrm{signal}}$ is the number of signal events
predicted by the \sherpa MC simulation,  with selection efficiencies extracted
from data.
The measured cross-section $\sigma_{\mathrm{data}}$ is
derived from the signal strength by multiplying it by the \sherpa MC
cross-section prediction $\sigma_{\mathrm{MC}}$ in the fiducial
region.

The signal strength is extracted using a likelihood fit over the
centrality of the $Z\gamma$ two-body system, $\cZg$ (see
Equation~(\ref{eqn:centrality}) and Figure~\ref{fig:templ_SR_CR}), which provides good discrimination
between the EWK and QCD production processes. Probability density
functions are built~\cite{Cranmer:1456844} from binned
histograms of $\cZg$ distributions (referred to as templates) using MC
events for signal and each of the backgrounds described in
Section~\ref{sec:background}.  The interference between the EWK and
QCD induced processes is not included in the probability density
functions but rather taken as
an uncertainty ($\sim7\%$ of the signal yield, determined
with \sherpa MC samples).

An extended likelihood is built from the product of four likelihoods
corresponding to $\cZg$ distributions in the SR and CR for the electron channel and the SR and  CR
for the muon channel. The inclusion of the CR likelihoods in the fit (where
the EWK signal process is suppressed) provides a strong constraint on
the QCD process normalization in the SR. 
The QCD normalization 
 is introduced in the likelihood as a single parameter for both the CR and the SR. It is
 treated as an unconstrained 
nuisance parameter and mainly determined by the data in the CR, where
events from the QCD process dominate. The normalizations
and shapes of the other backgrounds are taken from MC predictions and can vary within the uncertainties reported in
Section~\ref{sec:background}.

The signal strength for the EWK production, $\mu_{\mathrm{EWK}}$, and its
uncertainty are determined with a profile-likelihood-ratio test
statistic~\cite{Cowan:2011epjc}.  
Systematic uncertainties 
in the input templates are handled using nuisance parameters corresponding to
each systematic effect, which are assumed to have Gaussian
distributions with standard deviation $\sigma$ equal to the systematic uncertainty in the
parameter in question.
The profile of the negative log-likelihood ratio of the signal
strength $\mu_{\mathrm{EWK}}$ is shown in Figure~\ref{fig:logratio}.

\begin{figure}[h!]
  \begin{center}
 \includegraphics[width=0.60\columnwidth]{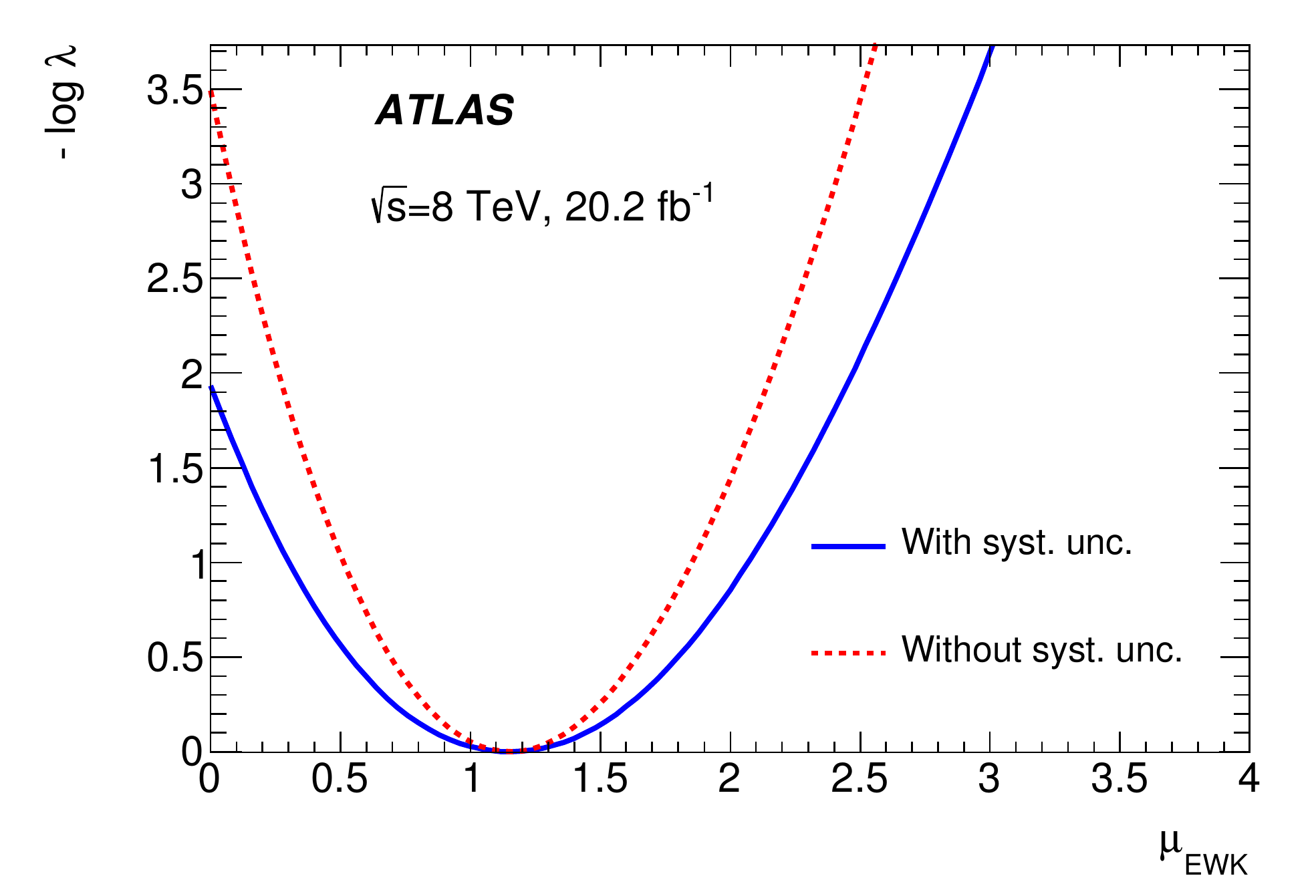}       
     \caption{The profile of the negative log-likelihood ratio $\lambda(\mu_{\mathrm{EWK}})$ of the
         signal strength $\mu_{\mathrm{EWK}}$ of the $\cZg$ template. The observed
         result is shown by the solid curve, while the dashed curve
         shows the results with only the statistical uncertainty included.
         The observed significance of the
         measurement is given by $\sqrt{-2\log\lambda(0)}$ and equals 2.0$\sigma$ in this case.}
    \label{fig:logratio}
  \end{center}
\end{figure}

From the best-fit value of $\mu_{\mathrm{EWK}}$, the observed \Zgjj EWK
production fiducial cross-section in the SR defined in Table~\ref{tab:selection_lep}, is
found to be:
\begin{align*}
  \sigma^{\mathrm{EWK}}_{Z\gamma jj} &= 1.1 \pm 0.5\ \text{(stat)} \pm 0.4\ \text{(syst)}\  \text{fb} = 1.1 \pm 0.6 \ \text{fb}.
\end{align*}
while the SM NLO prediction~\cite{bib:ZyNLO} from \vbfnlo, after applying the
\apart correction described above, and with uncertainties calculated as described in Section~\ref{sec:simulation}, is 
\begin{align*}
  \sigma^{\vbfnlo, \mathrm{EWK}}_{Z\gamma jj} &= 0.94 \pm 0.09\  \text{fb}.
\end{align*}
The significance of the observed EWK production signal is 2.0$\sigma$
(1.8$\sigma$ expected), not large enough to claim an observation of
this process.
The measured 95\% confidence level (CL) cross-section upper limit
obtained with the $CL_S$ technique~\cite{Cowan:2011epjc} is 2.2 fb.
A breakdown of the uncertainties of the $\sigma^{\mathrm{EWK}}_{Z\gamma jj}$ cross-section measurement,
as well as the total (EWK+QCD) cross-sections discussed in the next section,
is given in Table~\ref{tab:systXS}.

\begin{table}
 \caption{Summary of the main relative uncertainties in cross-section measurements presented in
 this paper.} \label{tab:systXS}
\begin{center}
\begin{tabular}{lrrr}
\hline
  Source of & \multicolumn{1}{r}{EWK [\%]} & \multicolumn{2}{r}{Total (EWK+QCD) [\%]} \\
  uncertainty &  & \multicolumn{1}{r}{SR} & \multicolumn{1}{r}{CR} \\
\hline
  Statistical & 40 & 9 & 4\\
\hline
  Jet energy scale & 36 & 9 & 4\\
  Theory & 10 & 5 & 4\\
  All other & 8 & 5 & 6\\
  Total systematic & 38 & 11 & 8\\
\hline
\end{tabular}
\end{center}
\end{table}

The statistical uncertainty of the measurement of $\sigma^{\mathrm{EWK}}_{Z\gamma jj}$ is
40\%. The 38\% systematic uncertainty
is dominated by the 36\% jet energy scale (JES) uncertainty
contribution.  In particular, the contribution of the uncertainty in
the $\eta$-intercalibration method of the JES is quite large, since
events at low $\cZg$ tend to have jets with high rapidity. This
uncertainty strongly affects both the normalization and the shape of
the EWK and QCD $\cZg$ distribution.  The impact of the JES
uncertainty in the shape of the $\cZg$ distributions increases the
normalization-only uncertainty by about 40\%, and affects the SR and
CR uncertainties by different amounts.  As a consequence, the
constraint from CR data is not as effective in reducing the JES
uncertainty as it is for other systematic uncertainties. Nonetheless, the use of
the CR data reduces the total systematic uncertainty of the EWK
cross-section measurement in the SR from about 60\% to 38\%.

The second largest contribution to the systematic uncertainty is from the theory uncertainty
(\sherpa modelling of the $Z\gamma jj$ production processes and interference between the QCD and EWK processes) and amounts to 10\%,  
while all other contributions combined (photon and lepton identification, reconstruction,
isolation and energy scale, and uncertainty from the $Z+$jet background estimate) are around
8\%. 

The post-fit QCD production normalization 
 is found to be in agreement
with the \sherpa predictions within one standard deviation. 
Cross-sections extracted separately in the
electron and muon channels are also compatible within their statistical
uncertainty.

\subsection{Total $Z\gamma jj$ fiducial cross-section measurements}

The total \Zgjj production (QCD+EWK) cross-section in the SR and CR is
extracted from data with a slight modification of the template fit
method described in Section~\ref{sec:XSewk}. In this case both the EWK
and QCD \Zgjj production are considered as signal. As a consequence a
unique signal strength $\mu_{Z\gamma jj}$ for the total QCD and EWK
production is used, and hence the relative ratio of QCD and EWK
contributions are fixed to the \sherpa MC predicted value.  The SR and
CR phase-space regions are treated independently and
$\mu_{Z\gamma jj}$ is not assumed to be the same in the two
regions. As before, $\cZg$ templates are taken from the separate
\sherpa MC samples of the EWK and QCD production.

EWK--QCD interference is negligible with respect to the total production and is not considered.

From the best-fit value of $\mu_{Z\gamma jj}$, the measured total
\Zgjj production cross-section in the SR and CR is found to be:
 \begin{align*}
  \sigma^{\mathrm{SR}}_{Z\gamma jj} &= 3.4 \pm 0.3 \ \text{(stat)} \pm 0.4 \ \text{(syst}) \ \text{fb} = 3.4 \pm 0.5 \ \text{fb}, \\
  \sigma^{\mathrm{CR}}_{Z\gamma jj} &= 21.9 \pm 0.9 \ \text{(stat)} \pm 1.8 \ \text{(syst)} \ \text{fb} = 21.9 \pm 2.0 \ \text{fb}. 
\end{align*}

The cross-sections measured separately in the
electron and muon channels are compatible within their statistical
uncertainty.

In both measurements, the systematic uncertainty (11\% and 8\% in the SR
and CR respectively) dominates over the statistical uncertainty (9\% and 4\% in the SR
and CR respectively).
A summary of the main sources of systematic uncertainty is given in
Table~\ref{tab:systXS}. 

The total \Zgjj cross-section predictions at QCD NLO~\cite{bib:ZyNLO} are derived adding
the individual \vbfnlo\ QCD and EWK production predictions with uncertainties 
calculated as described in Section~\ref{sec:simulation}. They are
found to be: 
\begin{align*}
  \sigma^{\vbfnlo, \text{SR}}_{Z\gamma jj} &= 4.0 \pm 0.4\  \text{fb}, \\
  \sigma^{\vbfnlo, \text{CR}}_{Z\gamma jj} &= 22.9 \pm 1.9\  \text{fb}.     
\end{align*}
These predictions include \apart
corrections derived from \sherpa MC samples reported before for EWK production
as well as \apart corrections for QCD production, which are 0.76 and 0.82 in the CR and SR
respectively, with negligible statistical uncertainty.

A summary of the \Zgjj production cross-section measurements in the charged-lepton channel is
given in Table~\ref{tab:cs_summary}. 
In both the SR and CR, and for both electroweak \Zgjj and total \Zgjj production,
the \vbfnlo\ MC predictions describe the data well. 

\begin{table}[h!]
  \caption{Summary of \Zgjj production cross-section measurements in the search and control
     regions for the charged-lepton channel.} 
  \centering
  \begin{tabular}{lccr@{$\,\pm\,$}c@{$\,\pm\,$}lr@{$\,\pm\,$}l}
    \hline
Channel & Phase-space  & Process & \multicolumn{3}{c}{Measured}           & \multicolumn{2}{c}{Predicted} \\
        & region       & type    & \multicolumn{3}{c}{cross-section [fb]} & \multicolumn{2}{c}{cross-section [fb]} \\
    \hline
$Z(\ell^+\ell^-)\gamma jj$ & Search region & EWK      & 1.1 & 0.5\ \text{(stat)}  & 0.4\ \text{(syst)}  & 0.94 &  0.09 \\
$Z(\ell^+\ell^-)\gamma jj$ &Search region & EWK+QCD   & 3.4 & 0.3 \ \text{(stat)} & 0.4 \ \text{(syst)} & 4.0  &  0.4  \\
$Z(\ell^+\ell^-)\gamma jj$ &Control region & EWK+QCD  & 21.9  & 0.9 \ \text{(stat)}   & 1.8 \ \text{(syst)}   & 22.9   &  1.9    \\
    \hline
  \end{tabular}
  \label{tab:cs_summary}
\end{table}

\section{Limits on quartic gauge-boson couplings}
\label{sec:aqgc}

Since QGCs contribute to VBS processes in the SM,
the \Zgjj EWK production, which contains VBS processes,
can be used to probe new physics via anomalies in the extracted coupling values.
\Zgjj EWK production in the SM has no
contributions at the tree level from the neutral QGCs of $ZZZ\gamma$,
$ZZ\gamma\gamma$ and $Z\gamma\gamma\gamma$ which consist of only the vector bosons with zero electric charges,
but does contain the charged QGC $WWZ\gamma$ vertex with the presence of charged vector bosons in the couplings.
New physics beyond the SM could induce
charged and neutral anomalous QGCs, enhancing the \Zgjj
electroweak production cross-section and modifying the kinematic
distributions of the final-state bosons.

In this section, events selected in the neutrino channel and in the
``aQGC region'' for the charged-lepton channel are used to probe aQGC
contributions to the \Zgjj EWK production, whose relative size is expected to
increase with the photon transverse energy. Optimization studies
performed on MC samples show that the selection of events with
$E_{\mathrm{T}}^{\gamma}>250$~\GeV ($E^{\mathrm{T}}_{\gamma}>150$~\GeV) for the charged-lepton 
(neutrino) channel maximizes the sensitivity to aQGC
contributions for the available amount of data.
Figures~\ref{fig:photon_pt_fm0} and ~\ref{fig:nunu_pt_fm0} show the photon
transverse energy distribution in the charged-lepton and neutrino channel, respectively,
for the data as well as the expected signal and backgrounds, along with one 
possible aQGC contribution. 

\begin{figure}[h!]
  \begin{center}
    \includegraphics[width=0.60\columnwidth]{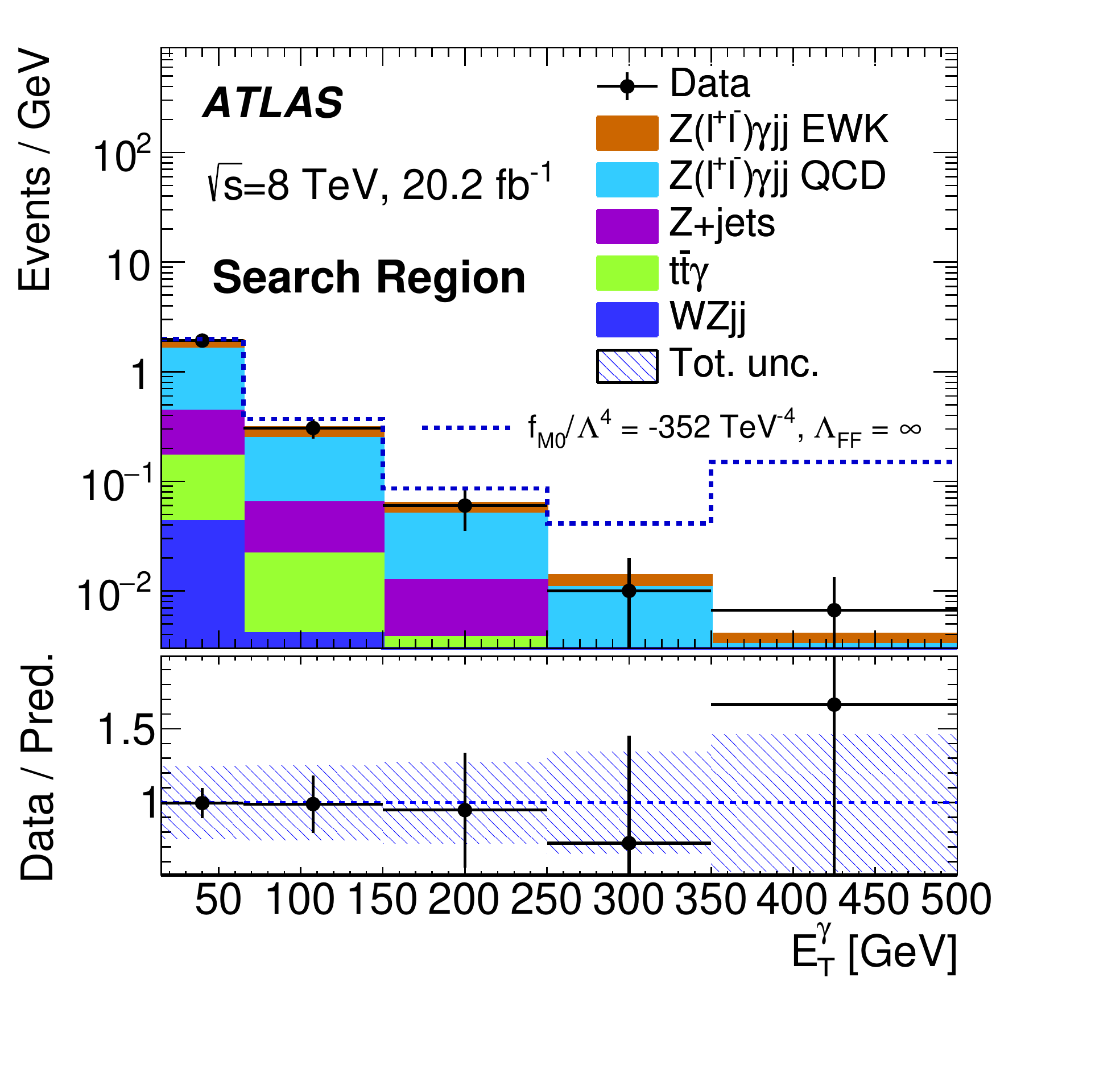}
    \caption{Distributions of the photon transverse energy for the electron and muon channels
      added together in the search region,
      for the data (black points), and for the signal process and
      various background components (coloured templates) before any
      fit is done. 
      The sum of the signal prediction with one non-zero aQGC parameter and the
      various backgrounds discussed in the text is also shown (blue dashed line).  
      The ratio between data and the sum of all pre-fit expected contributions (``Pred.'')
      is shown below the histogram. The hatched blue band shows the
      systematic and statistical uncertainty added in quadrature (``Tot. unc.'') in the signal and background
      prediction, 
      while the error bars on the data points represent the statistical uncertainty of the data set. 
      The number of events in each bin is divided by the bin width. The last bin also
      includes events beyond the range shown. } 
    \label{fig:photon_pt_fm0}
  \end{center}
\end{figure}

\begin{figure}[h!]
  \begin{center}
    \includegraphics[width=0.60\columnwidth]{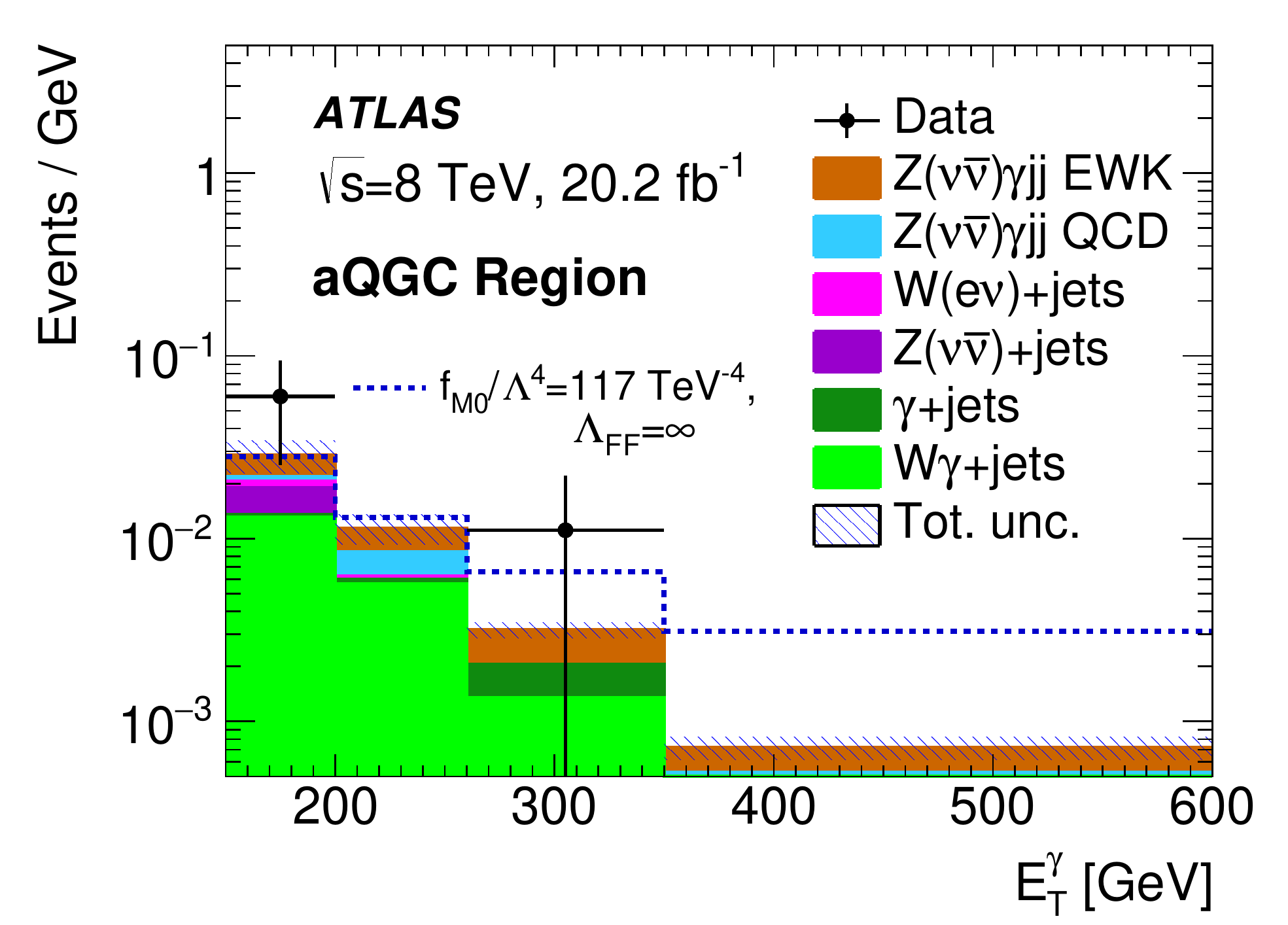}
    \caption{Distributions of the photon transverse energy for the neutrino channel
      in the aQGC region,
      for the data (black points), and for the signal process and
      various background components (coloured templates). 
      The sum of the signal prediction with one non-zero aQGC parameter and the
      various backgrounds discussed in the text is also shown (blue dashed line).  
      The hatched blue band shows the systematic and statistical
      uncertainty added in quadrature (``Tot. unc.'') in the signal
      and background prediction, while the error bars on the data
      points represent the statistical uncertainty of the data set.
      The number of events in each bin is divided by the bin
      width. The last bin also includes events beyond the range
      shown. }
    \label{fig:nunu_pt_fm0}
  \end{center}
\end{figure}

\subsection{Fiducial EWK production cross-section limits in high-$E_{\mathrm{T}}^{\gamma}$ regions}
\label{sec:highEtxs}

In this section, limits on the fiducial cross-section of  \Zgjj EWK
production in the charged-lepton and neutrino channels
in the high-$E_{\mathrm{T}}^{\gamma}$ energy region
dedicated to aQGC studies are reported.  The 
fiducial phase-space region definition for the neutrino (charged-lepton)
channel is provided in Table~\ref{tab:selection_nu}
(Table~\ref{tab:selection_lep}).
These selections follow closely the experimental
selection reported in Section~\ref{sec:nu_sel}, with MC
particle-level objects defined as in Section~\ref{sec:xsec}.

\begin{table}[h!]
  \centering
  \caption{Neutrino channel aQGC region definition at particle
    (parton) level. If there are more than two jets at particle level,
    the two highest transverse momentum ones are considered.}
  \begin{tabular}{lcc}
    \hline
Objects & Particle- (Parton-) level selection \\
\hline 
Neutrinos & $E_{\mathrm{T}}^{\nu\bar{\nu}} > 100$~\GeV \\
\hline
Photon (kinematics)& $E_{\mathrm{T}}^{\gamma}>150$~\GeV, $|\eta^{\gamma}|<2.37$ \\
 & $\Delta R(\ell, \gamma)>0.4$ \\
\hline
Photon (isolation) & $E_{\mathrm{T}}^{\mathrm{iso}} < 0.5 \cdot E_{\mathrm{T}}^{\gamma}$ \\
\hline
Generator-level jets (Outgoing quarks) & At least two jets (quarks) \\
($pp \rightarrow Z \gamma qq$) & $E_{\mathrm{T}}^{j(q)}>30$ \GeV, $|\eta^{ j(q)}|<4.5$ \\
 & $\Delta R(\gamma, j(q))>0.4$ \\
\hline
Event kinematic & $|\Delta\phi(E_{\mathrm{T}}^{\nu\bar{\nu}},\gamma jj(qq))|>\frac{3\pi}{4}$ \\
selection & $|\Delta\phi(E_{\mathrm{T}}^{\nu\bar{\nu}},\gamma)|>\frac{\pi}{2}$ \\
 & $|\Delta\phi(E_{\mathrm{T}}^{\nu\bar{\nu}},j(q))|>1$\\ 
 & $E_{\mathrm{T}}^{\gamma}>150$ \GeV \\
 &  $|\Delta y_{jj(qq)}|>2.5$  \\
 &  $\zeta_{\gamma}$ < 0.3  \\
 &  $p_{\mathrm{T}}^{\mathrm{balance}}<$ 0.1 \\
 &  $m_{jj(qq)}>600$ \GeV \\
\hline
  \end{tabular}
  \label{tab:selection_nu}
\end{table}

For both channels, cross-sections are extracted from data
using a log-likelihood fit method, where the likelihood function is
constructed from a Poisson distribution for the expected number of signal and background events
in the aQGC region and
Gaussian constraints for nuisance parameters associated with systematic uncertainties.
These systematic uncertainties are summarized in Table~\ref{tab:systQGC}.

The \Zgjj EWK cross-sections predicted by the SM in
these high-$E_{\mathrm{T}}^{\gamma}$ fiducial regions are very small: 
($71\pm7$) ab and ($17\pm2$) ab for the neutrino and charged-lepton
channels, respectively. Therefore, with 20.2 fb$^{-1}$ of data only cross-section
upper limits are extracted.

Using the \SHERPA MC simulation, reconstruction efficiencies, defined as the number
of reconstructed MC events passing the event selection divided by the number of generated events at particle level within the
fiducial region, are found to be 
$C^{\nu\bar{\nu}\gamma jj}_{\mathrm{EWK}} = 0.54\pm0.03$ and
$C^{\ell^+\ell^-\gamma jj}_{\mathrm{EWK}} = 0.56\pm0.03$. These
factors account for photon and lepton reconstruction, identification,
isolation, trigger and jet identification efficiencies and related
systematic uncertainties. 

The numbers of observed events for the neutrino and charged-lepton
channels are four and two, respectively. The 95\% CL cross-section
upper limits obtained with the $CL_S$ technique are 1.06 fb (0.99 fb
expected) in the neutrino channel and 1.03 fb (1.01 fb expected) in
the charged-lepton channel.

\subsection{Extracting confidence intervals on anomalous quartic gauge-boson couplings.}
\label{sec:aqgcLimits}
An effective field theory (EFT)~\cite{bib:eft} with higher-dimensional
operators~\cite{bib:eboli-dim8} is adopted to parameterize the
anomalous couplings.  These operators are in the linear Higgs-doublet
representation~\cite{bib:eboli-dim8, bib:snowmass_ewk}.

A parity-conserving EFT Lagrangian is constructed based on the
hypothesis that the observed Higgs boson belongs to an
$\mathrm{SU(2)}_{\mathrm{L}}$ doublet~\cite{bib:eboli-dim8} and has the form

\begin{equation}
\label{eq:lagEFT}
{\cal L} = {\cal L}^{\mathrm{SM}} + \sum_{i}\frac{c_{i}}{\Lambda^{2}}{\cal O}_{i} + \sum_{j}\frac{f_{j}}{\Lambda^{4}}{\cal O}_{j}.
\end{equation}
The first term represents the SM lagrangian, while higher-order terms represent new physics inducing anomalous gauge couplings, 
such as the dimension-6 term (second term) and dimension-8 term (third term) with all components of each term summed up according to the dimension of the new physics scale~$\Lambda$~\cite{bib:snowmass_ewk,bib:eft}.
Out of these higher-order operators, the dimension-8 ones are the lowest-dimension operators
inducing only quartic gauge-boson couplings without triple gauge-boson
vertices.  The dimension-8 operators with coeffcients $f_j$ are sub-categorized in
$f_{T,x}$ operators, containing only the field-strength tensor, and
$f_{M,x}$ operators, containing both the Higgs ${\mathrm{SU(2)}}_\mathrm{L}$  doublet
derivatives and the field strength.
The charged anomalous couplings of $WWZ\gamma$ can be induced only by the
$f_{M,x}$ ($x$=0--7) and $f_{T,x}$ ($x$=0--7) operators.
The neutral aQGCs of $ZZZ\gamma$ and $ZZ\gamma\gamma$ can be induced by the
$f_{M,x}$ ($x$=0--7) and $f_{T,x}$ ($x$=0--9) operators
while $Z\gamma\gamma\gamma$ QGC can be modified only by
$f_{T,x}$ ($x$=0--9) operators.

Only a subset of these dimension-8 operators are probed here with
events from \Zgjj EWK production: $f_{T0}/\Lambda^{4}$, as a
representative operator of $f_{T1}/\Lambda^4$ and $f_{T2}/\Lambda^4$;
$f_{T8}/\Lambda^4$ and $f_{T9}/\Lambda^4$, the two unique operators
which can be probed only via neutral QGC vertices;
and $f_{M0}/\Lambda^4$,
$f_{M1}/\Lambda^4$, $f_{M2}/\Lambda^4$ and $f_{M3}/\Lambda^4$, which
allow comparisons and possible combinations with results from other
experiments.

Form factors (FF) must be introduced to restore unitarity at very high
parton centre-of-mass energy $\sqrt{\hat{s}}$:
$f_i(\hat{s})=f_{i}/(1+\hat{s}/\Lambda_{\mathrm{FF}}^2)^n$,
where $\Lambda_{\mathrm{FF}}$ is the parameter chosen to prevent unitarity
violation up to $\sqrt{\hat{s}} = 8$ \TeV~and the FF exponent is set to $n=2$.

Data in the aQGC regions, defined in Tables~\ref{tab:selection_lep}
and~\ref{tab:selection_nu} for the charged-lepton and neutrino channels
respectively, are used. These phase-space regions are determined
by optimizing the expected confidence intervals on the dimension-8 operator
coefficients in the individual channels.

MC \Zgjj EWK events with non-zero EFT dimension-8 operator
coefficients are generated by \madgraph5.1~\cite{Madgraph} at LO,
interfaced with \pythia6~\cite{Sjostrand:2006za} for parton showering, and
processed through the ATLAS detector full simulation, for both the
charged-lepton and the neutrino channels. One MC sample is generated
for each of the dimension-8 operators to study the aQGC event
selection efficiencies. The difference with respect to the selection
efficiency calculated from SM samples modelled with \sherpa is taken
as an additional uncertainty of 20\%(10\%) for the charged-lepton
(neutrino) channel analysis. The event yield from background physics
processes sensitive to aQGC probed here are kept fixed to SM
predictions.

The aQGC cross-sections are calculated at NLO precision in
$\alpha_{\mathrm{s}}$~\cite{bib:ZyNLO} using the \vbfnlo{} event generator 
and parameterized as a function of the dimension-8 operator anomalous
coupling values. For each operator, the cross-sections are calculated
for five aQGC coupling values. A parabola is then fit to parameterize
the cross-section as a function of the operator coefficient.

For both channels, a one-dimensional profile-likelihood fit is used to
derive the expected and observed 95\% CL intervals of each dimension-8
operator coefficient, with all the other operator coefficients set to
zero.  The neutrino channel provides the best expected intervals for all
operators. However, they are improved by 10--30\% when the results from
both channels are combined.  The channel combination is obtained
taking into account correlations between systematic uncertainties.  In
particular, the systematic uncertainties of the photon identification
efficiency, uncertainties due to the jet energy scale and resolution, PDF
uncertainties, uncertainties in parton-shower modelling, and the
systematics of QCD renormalization and factorization scales are
considered as correlated. Among these systematic uncertainties,
the QCD renormalization and factorization scale uncertainties are the
leading ones, and are 8.0\% (7.7\%)
in the charged-lepton (neutrino) channel.

Table~\ref{tab:observedExpected1DLimitsAQGC_comb} shows the expected
and observed 95\% CL intervals derived with the $\Lambda_{\mathrm{FF}}$
set to the maximum allowed value to preserve unitarity, according to
the \vbfnlo{} calculation. The expected and observed 95\% CL intervals derived without respecting
the unitarity bounds ($n=0$) and without other unitarization
restrictions are also reported  in
Table~\ref{tab:observedExpected1DLimitsAQGC_comb}.

\begin{table}
\caption{Measured and expected one-dimensional 95\% confidence level intervals
  on aQGC parameters (in the \vbfnlo{} formalism) 
  using the combination of all \Zgjj channels (charged-lepton and neutrino). 
  The FF exponent $n=0$ entries correspond to an infinite FF scale and therefore result in
  non-unitarized 95\% CL inntervals. FF exponent $n=2$ confidence intervals preserve unitarity with individual form-factor 
  scales as shown in the last column for each dimension-8 operator.
  The maximum allowed form-factor values $\Lambda_{\mathrm{FF}}$ are chosen, according to the unitarity bounds
  calculated by \vbfnlo{} and are also reported in this table. The latest NLO cross-section prediction of \vbfnlo{} is used in the aQGC parameterization.~\cite{bib:ZyNLO}} 
\begin{center}
\begin{tabular}{ccccc}
        \hline
              & 95\% CL intervals  & Measured [\TeV{}$^{-4}$] & Expected [\TeV{}$^{-4}$] & $\Lambda_{\mathrm{FF}}$ [\TeV{}] \\ \hline
\multirow{7}{*}{$n=0$} & $f_{T9}/\Lambda^4$ & $[-4.1, 4.2]\times10^3$  & $[-2.9, 3.0]\times10^3$ & \\
              & $f_{T8}/\Lambda^4$ & $[-1.9, 2.1]\times10^3$  & $[-1.2, 1.7]\times10^3$ &  \\
              & $f_{T0}/\Lambda^4$ & $[-1.9, 1.6]\times10^1$  & $[-1.6, 1.3]\times10^1$ & \\
              & $f_{M0}/\Lambda^4$ & $[-1.6, 1.8]\times10^2$  & $[-1.4, 1.5]\times10^2$ & \\
              & $f_{M1}/\Lambda^4$ & $[-3.5, 3.4]\times10^2$  & $[-3.0, 2.9]\times10^2$ & \\
              & $f_{M2}/\Lambda^4$ & $[-8.9, 8.9]\times10^2$  & $[-7.5, 7.5]\times10^2$ & \\
              & $f_{M3}/\Lambda^4$ & $[-1.7, 1.7]\times10^3$  & $[-1.4, 1.4]\times10^3$ & \\ \hline
\multirow{5}{*}{$n=2$} & $f_{T9}/\Lambda^4$ & $[-6.9, 6.9]\times10^4$  & $[-5.4, 5.3]\times10^4$ & 0.7\\
              & $f_{T8}/\Lambda^4$ & $[-3.4, 3.3]\times10^4$  & $[-2.6, 2.5]\times10^4$ & 0.7 \\
              & $f_{T0}/\Lambda^4$ & $[-7.2, 6.1]\times10^1$  & $[-6.1, 5.0]\times10^1$ & 1.7 \\
              & $f_{M0}/\Lambda^4$ & $[-1.0, 1.0]\times10^3$  & $[-8.8, 8.8]\times10^2$ & 1.0\\
              & $f_{M1}/\Lambda^4$ & $[-1.6, 1.7]\times10^3$  & $[-1.4, 1.4]\times10^3$ & 1.2 \\
              & $f_{M2}/\Lambda^4$ & $[-1.1, 1.1]\times10^4$  & $[-9.2, 9.6]\times10^3$ & 0.7 \\
              & $f_{M3}/\Lambda^4$ & $[-1.6, 1.6]\times10^4$  & $[-1.4, 1.3]\times10^4$ & 0.8 \\ \hline
\end{tabular}
\end{center}
\label{tab:observedExpected1DLimitsAQGC_comb}
\end{table}

\clearpage

\section{Conclusions}
\label{sec:conclusions}

Studies of electroweak production of a $Z$ boson and a photon in
association with a high-mass dijet system, with the $Z$ boson decaying
into a pair of either electrons, muons, or neutrinos, have been
performed using 20.2 fb$^{-1}$ of proton--proton collision data at $\sqrt{s}=8$~\TeV
collected by the ATLAS experiment at the LHC.

In the charged-lepton channel, the \Zgjj EWK production cross-section
in a fiducial region with a signal purity of about 18\% is
found to be:
\[  \sigma^{\mathrm{EWK}}_{Z\gamma jj} = 1.1 \pm 0.6 \ \text{fb}, \]
which is consistent with the NLO SM prediction from \vbfnlo, 
and corresponds to a significance of $2\sigma$. 
The total \Zgjj production cross-section is also measured in the
same fiducial region and in a phase-space region dominated by QCD
production and both are found to be in good agreement with the
NLO SM predictions from \vbfnlo.

Events with high-$E_{\mathrm{T}}$ photons in both the charged-lepton 
and neutrino decay modes of the $Z$ boson are used to extract confidence intervals 
on seven different aQGC parameters modelled by dimension-8 operators of
an effective field theory.  In both channels, 95\% CL upper limits
on the SM \Zgjj EWK production cross-sections are placed in these
high-$E_{\mathrm{T}}$ photon phase-space regions. The 95\% CL intervals on the aQGC
parameters are competitive or even more stringent than previous
constraints obtained with events with \Zgjj and different final states.

\section*{Acknowledgements}


We thank CERN for the very successful operation of the LHC, as well as the
support staff from our institutions without whom ATLAS could not be
operated efficiently.

We acknowledge the support of ANPCyT, Argentina; YerPhI, Armenia; ARC, Australia; BMWFW and FWF, Austria; ANAS, Azerbaijan; SSTC, Belarus; CNPq and FAPESP, Brazil; NSERC, NRC and CFI, Canada; CERN; CONICYT, Chile; CAS, MOST and NSFC, China; COLCIENCIAS, Colombia; MSMT CR, MPO CR and VSC CR, Czech Republic; DNRF and DNSRC, Denmark; IN2P3-CNRS, CEA-DSM/IRFU, France; SRNSF, Georgia; BMBF, HGF, and MPG, Germany; GSRT, Greece; RGC, Hong Kong SAR, China; ISF, I-CORE and Benoziyo Center, Israel; INFN, Italy; MEXT and JSPS, Japan; CNRST, Morocco; NWO, Netherlands; RCN, Norway; MNiSW and NCN, Poland; FCT, Portugal; MNE/IFA, Romania; MES of Russia and NRC KI, Russian Federation; JINR; MESTD, Serbia; MSSR, Slovakia; ARRS and MIZ\v{S}, Slovenia; DST/NRF, South Africa; MINECO, Spain; SRC and Wallenberg Foundation, Sweden; SERI, SNSF and Cantons of Bern and Geneva, Switzerland; MOST, Taiwan; TAEK, Turkey; STFC, United Kingdom; DOE and NSF, United States of America. In addition, individual groups and members have received support from BCKDF, the Canada Council, CANARIE, CRC, Compute Canada, FQRNT, and the Ontario Innovation Trust, Canada; EPLANET, ERC, ERDF, FP7, Horizon 2020 and Marie Sk{\l}odowska-Curie Actions, European Union; Investissements d'Avenir Labex and Idex, ANR, R{\'e}gion Auvergne and Fondation Partager le Savoir, France; DFG and AvH Foundation, Germany; Herakleitos, Thales and Aristeia programmes co-financed by EU-ESF and the Greek NSRF; BSF, GIF and Minerva, Israel; BRF, Norway; CERCA Programme Generalitat de Catalunya, Generalitat Valenciana, Spain; the Royal Society and Leverhulme Trust, United Kingdom.

The crucial computing support from all WLCG partners is acknowledged gratefully, in particular from CERN, the ATLAS Tier-1 facilities at TRIUMF (Canada), NDGF (Denmark, Norway, Sweden), CC-IN2P3 (France), KIT/GridKA (Germany), INFN-CNAF (Italy), NL-T1 (Netherlands), PIC (Spain), ASGC (Taiwan), RAL (UK) and BNL (USA), the Tier-2 facilities worldwide and large non-WLCG resource providers. Major contributors of computing resources are listed in Ref.~\cite{ATL-GEN-PUB-2016-002}.

\clearpage

\printbibliography

\newpage
\begin{flushleft}
{\Large The ATLAS Collaboration}

\bigskip

M.~Aaboud$^\textrm{\scriptsize 137d}$,
G.~Aad$^\textrm{\scriptsize 88}$,
B.~Abbott$^\textrm{\scriptsize 115}$,
J.~Abdallah$^\textrm{\scriptsize 8}$,
O.~Abdinov$^\textrm{\scriptsize 12}$$^{,*}$,
B.~Abeloos$^\textrm{\scriptsize 119}$,
S.H.~Abidi$^\textrm{\scriptsize 161}$,
O.S.~AbouZeid$^\textrm{\scriptsize 139}$,
N.L.~Abraham$^\textrm{\scriptsize 151}$,
H.~Abramowicz$^\textrm{\scriptsize 155}$
% ...
\bigskip
\\
% ...
$^{8}$ Department of Physics, The University of Texas at Arlington, Arlington TX, United States of America\\
% ...
$^{12}$ Institute of Physics, Azerbaijan Academy of Sciences, Baku, Azerbaijan\\
% ...
$^{88}$ CPPM, Aix-Marseille Universit{\'e} and CNRS/IN2P3, Marseille, France\\
% ...
$^{115}$ Homer L. Dodge Department of Physics and Astronomy, University of Oklahoma, Norman OK, United States of America\\
% ...
$^{119}$ LAL, Univ. Paris-Sud, CNRS/IN2P3, Universit{\'e} Paris-Saclay, Orsay, France\\
% ...
$^{137}$ $^{(a)}$ Facult{\'e} des Sciences Ain Chock, R{\'e}seau Universitaire de Physique des Hautes Energies - Universit{\'e} Hassan II, Casablanca; $^{(b)}$ Centre National de l'Energie des Sciences Techniques Nucleaires, Rabat; $^{(c)}$ Facult{\'e} des Sciences Semlalia, Universit{\'e} Cadi Ayyad, LPHEA-Marrakech; $^{(d)}$ Facult{\'e} des Sciences, Universit{\'e} Mohamed Premier and LPTPM, Oujda; $^{(e)}$ Facult{\'e} des sciences, Universit{\'e} Mohammed V, Rabat, Morocco\\
% ...
$^{139}$ Santa Cruz Institute for Particle Physics, University of California Santa Cruz, Santa Cruz CA, United States of America\\
% ...
$^{151}$ Department of Physics and Astronomy, University of Sussex, Brighton, United Kingdom\\
% ...
$^{155}$ Raymond and Beverly Sackler School of Physics and Astronomy, Tel Aviv University, Tel Aviv, Israel\\
% ...
$^{161}$ Department of Physics, University of Toronto, Toronto ON, Canada\\
% ...
$^{a}$ Also at Department of Physics, King's College London, London, United Kingdom\\
$^{b}$ Also at Institute of Physics, Azerbaijan Academy of Sciences, Baku, Azerbaijan\\
$^{c}$ Also at Novosibirsk State University, Novosibirsk, Russia\\
$^{d}$ Also at TRIUMF, Vancouver BC, Canada\\
$^{e}$ Also at Department of Physics {\&} Astronomy, University of Louisville, Louisville, KY, United States of America\\
% ...
$^{*}$ Deceased
\end{flushleft}


\end{document}